\documentclass[
superscriptaddress,
nofootinbib,
 amsmath,amssymb,
 aps,
]{revtex4-2}

\usepackage{graphicx}
\usepackage{dcolumn}
\usepackage{bm}
\usepackage{hyperref}
\usepackage[mathlines]{lineno}
\usepackage[dvipsnames]{xcolor}

\usepackage{comment}

\definecolor{mpl_blue}{HTML}{1F77B4}
\definecolor{mpl_orange}{HTML}{FF7F0E}
\definecolor{mpl_green}{HTML}{2CA02C}
\definecolor{mpl_red}{HTML}{D62728}

\newcommand{\diff}{\mathop{}\!\mathrm{d}}
\usepackage[normalem]{ulem}

\begin{document}


\title{The WDM time-frequency transform in gravitational-wave data analysis II:
Evolutionary spectrum and noise covariance}

\author{Aaron~D. Johnson}
\email{aaron.d.johnson@nasa.gov}
\affiliation{NASA Marshall Space Flight Center, Huntsville, AL 35812, USA}
\affiliation{Science and Technology Institute, Universities Space Research Association, Huntsville, AL 35805, USA}
\affiliation{TAPIR, California Institute of Technology, Pasadena, CA 91125, USA}%
\author{Katerina Chatziioannou}
\email{kchatziioannou@caltech.edu}
\affiliation{TAPIR, California Institute of Technology, Pasadena, CA 91125, USA}%
\affiliation{LIGO Laboratory, California Institute of Technology, Pasadena, CA 91125, USA}%
\author{Robert Rosati}
\email{robert.rosati@uah.edu}
\affiliation{NASA Marshall Space Flight Center, Huntsville, AL 35812, USA}
\affiliation{The University of Alabama in Huntsville, Huntsville, AL, 35805, USA}

\begin{abstract}
Noise cannot be treated as stationary over the timescales required for upcoming gravitational-wave detectors, minutes for ground-based and months for space-based detectors.
Nonstationarity leads to a non-diagonal frequency-domain noise covariance matrix, thus motivating a shift to the time-frequency domain.
We introduce \emph{evolutionary power spectra} for describing nonstationary noise in the Wilson-Daubechies-Meyer time-frequency basis.
Under the framework of oscillatory processes, we show that, to leading order in the ratio of the Meyer filter and the nonstationarity timescales, the evolutionary spectrum is a power-frequency distribution that can be estimated from a single noise realization.
We derive the corresponding noise covariance matrix and show that it is band tri-diagonal, like its stationary counterpart.
The exact covariance to all orders is positive semi-definite by construction and dense depending on the noise.
We further derive the optimal tiling of the time-frequency plane that minimizes the error of a diagonal covariance approximation and show that coarse-graining the plane suppresses correlations quadratically.
Using simulated data with a LISA Galactic foreground and a drifting powerlaw spectrum,
we show that the stationarity approximation leads to signal-to-noise ratio loss and severely biases inference.
For realistic LISA data, approximating the covariance as diagonal in the Wilson-Daubechies-Meyer domain is accurate for the narrow-band analysis of a single resolved Galactic binary, but insufficient for a broad-band analysis.
Finally, we provide mitigation strategies for the wraparound correlations induced across the data time edges whenever the spectrum differs between the start and the end of the data.
\end{abstract}

\maketitle

\section{Introduction}

Most current analyses of gravitational-wave data rely on the assumption that the stochastic component of the data is a stationary and Gaussian process.
The Gaussianity assumption determines the form of the likelihood function, while stationarity motivates frequency-domain analyses where the Fourier bins are uncorrelated and the noise covariance matrix is diagonal~\cite{LIGOScientific:2019hgc}.
The focus of this work is nonstationarity.

Both the nature of the stochastic process and the way in which stationarity fails depend on the detector.
For ground-based detectors at current sensitivity, the dominant stochastic process is instrumental noise which is approximately stationary on the timescale of transient compact binary signals, tens of milliseconds to 1-2 minutes. 
Longer-term nonstationarity complicates estimation of the noise power spectral density (PSD) directly from the detector data, motivating PSD-drift corrections~\cite{Zackay:2019kkv} or reranking of search candidates~\cite{Mozzon:2020gwa} to mitigate false positives.
Inference of the source properties, which can afford longer analysis times than the fast searches, instead parametrizes the noise PSD and infers it from the same data as the transient signal, thus easing the stationarity requirement from hours to minutes of data~\cite{Littenberg:2014oda,Plunkett:2022zmx,Gupta:2023jrn}.
In pulsar timing arrays, the dominant stochastic process is the signal itself: the gravitational-wave background emitted from an ensemble of supermassive black holes.
In that case, though the process itself is stationary, the finite observation time combined with the fact that the process has a steep red spectrum correlates Fourier bins and can bias the inferred spectral parameters~\cite{Crisostomi:2025vue}.

The upcoming space-based LISA detector will be dominated by different stochastic processes depending on the frequency range and data channel: instrumental noise, astrophysical foreground from unresolvable Galactic binaries, potential foregrounds from extragalactic sources depending on their astrophysical rates, and a cosmological background.
All these, plus the resolved signals, will be modeled and inferred simultaneously~\cite{Littenberg:2020bxy,Littenberg:2023xpl,Katz:2024oqg,Boileau:2021sni,Baghi:2023qnq,Rosati:2024lcs,Santini:2025iuj,Muratore:2023gxh,Aimen:2025zxn,Vajpeyi:2026pzh,Liu:2026yzk}.
While instrumental noise is also unlikely to be stationary, the Galactic foreground is demonstrably nonstationary, modulated at harmonics of $1/\mathrm{yr}$ as LISA sweeps the sky toward and away from the Galactic center~\cite{Digman:2022jmp,Buscicchio:2024wwm,Criswell:2024hfn,Buscicchio:2025zeb,Mathur:2026prt}.
Gaps in the data stream also induce nonstationarity and correlations between Fourier bins~\cite{Baghi:2019eqo,Burke:2025bun,Pearson:2025wfd,Burke:2026rjq}.
Proposed ways to deal with nonstationarity include segmenting the data when tests identify nonstationarity~\cite{Edwards:2020tlp} and moving analyses to the time-frequency domain with a short-time Fourier transform~\cite{Hindmarsh:2024ttn,Du:2025fes,Tenorio:2025gci}, or, the focus of this work, a Wilson-Daubechies transform~\cite{Cornish:2020odn,Vajpeyi:2026msr,Johnson:2026rrn}.

The critically-sampled Wilson-Daubechies transform is an orthogonal and complete mapping to the time-frequency domain~\cite{Wilson1987,daubechies1992ten}.
In gravitational-wave data analysis, it is most commonly paired with the Meyer filter~\cite{Meyer1992} which is compact in frequency~\cite{Necula_2012}, resulting in the Wilson-Daubechies-Meyer (WDM) basis.\footnote{While the Wilson-Daubechies construction forms the cornerstone of the transform's properties, orthogonality and completeness, it can be paired with any filter that satisfies certain conditions. For example a Gaussian-based filter might yield more compact time and frequency representations~\cite{Pearson:2026hgp}.}
Beyond addressing nonstationarity, in the LISA context the WDM transform has been employed for stellar-origin black hole binary inference assuming known noise~\cite{Digman:2022igm}, mitigating the impact of data gaps with Bayesian data augmentation~\cite{Pearson:2025wfd,Yan:2026nch}, and unmodeled transient searches~\cite{Tiwari:2026aea}.

This study is a companion to \citet{Johnson:2026rrn}, which offered a self-contained account of the theory of time-frequency methods, presented detailed derivations of the forward and inverse WDM transforms for series (such as the data) and matrices (such as the noise covariance), and derived the WDM likelihood. 
\citet{Johnson:2026rrn} presented generic equations for the noise covariance and posited the form of an \emph{evolutionary power spectrum} that promotes the standard noise PSD to a function of time, $S(t,f)$. 

In this paper, we present in detail the underlying theory, definition, and properties of the evolutionary spectrum.
Following the treatment of oscillatory processes by \citet{Priestley1981}, in Sec.~\ref{sec:evolutionary} we formally define the evolutionary PSD.
We show that for random processes with a nonstationarity timescale $B_x$ longer than the temporal extent of the filter $B_\phi$, $B_\phi \ll B_x$, the evolutionary spectrum can be interpreted as a physical distribution of power-frequency.
Its band-averaged form can be estimated from a process realization without additional information about the process properties.
In Sec.~\ref{sec:evolutionary-wdm}, we apply the evolutionary spectrum theory to the WDM transform, where the relevant filter is the Meyer filter, and derive equations for the WDM noise covariance matrix given a model for the evolutionary spectrum.
We show that to order $B_\phi/B_x$, the noise covariance matrix is block tri-diagonal, as is the case for purely stationary noise~\cite{Cornish:2025awt}.
Both its diagonal and off-diagonal elements are integrals of the spectrum over the Meyer filter.
Generic noise is handled in Sec.~\ref{sec:cov-generation} where we show how to calculate the, generically dense and not banded, noise covariance and simulate a noise realization given a known random process, albeit using knowledge about the process properties beyond those available from a single realization.
The evolutionary spectrum under the oscillatory-process theory of \citet{Priestley1981} is similar to the Gramian dynamical spectrum construction in \citet{Cornish:2026tjt}.
The correspondence between the two is further discussed in Sec.~\ref{sec:cov-generation}.

Switching to practical considerations, in Sec.~\ref{sec:optimal} we derive the optimal tiling of the WDM time-frequency plane such that the covariance is as diagonal as possible. 
As a diagonal WDM covariance is only achieved for purely stationary and white noise, the optimal tiling balances a small WDM cell time size $\Delta T$ (to hide nonstationarity) with a small frequency size $\Delta F$ (to hide the spectrum color), subject to the uncertainty principle, $\Delta F \Delta T=1/2$.
In Sec.~\ref{sec:coarse-graining} we explore time-frequency coarse-graining of the evolutionary PSD, similar to Welch estimates in the frequency domain.
We show that coarse-graining reduces cross-region correlations quadratically, at the expense of time and frequency resolution.

Finally, after introducing our inference methods in Sec.~\ref{sec:methods}, in Sec.~\ref{sec:results} we examine two inference simulations in the WDM domain: a sinusoidal signal in powerlaw noise with a drifting index and a LISA resolved Galactic binary in cyclostationary noise induced by unresolved binaries.
Inspecting the covariance structure of both, we show that the LISA foreground remains close to band tri-diagonal, while the drifting powerlaw exhibits large correlations that extend well beyond neighboring frequency bands.
We discuss a particular correlation-inducing effect: the WDM transform, like the Fourier transform, is subject to data wraparound so large correlations are concentrated around the time edges of the data.
We show that these correlations can severely bias inference for noise whose spectrum changes considerably between the start and end of the data, but they can be effectively mitigated with data excision or zero-padding.
Beyond time-edge correlations, applying the optimal tiling criterion to the LISA foreground, we find that a diagonal covariance is admissible across the narrow frequency band containing a single resolved Galactic binary, but not across the full bandwidth of a global fit.
A stationary likelihood severely biases inference of signal and noise, and reduces the recovered signal-to-noise ratio.
We examine four WDM likelihoods: (i) the exact covariance that is formally not accessible for real data, (ii) the tri-diagonal covariance that is valid to leading order in the noise nonstationarity, (iii) a diagonal approximation that restricts to the diagonal terms of the tri-diagonal covariance, and (iv) a diagonal approximation that further simplifies the diagonal terms.
WDM likelihoods recover broadly consistent noise and signal parameters with small shifts for a narrow-band single resolved binary.
We conclude in Sec.~\ref{sec:conclusions}.

\subsection{Summary of WDMI and conventions}
\label{sec:notation}

We adopt the conventions and notation of \citet{Johnson:2026rrn}, hereafter WDMI. 
In this section we collect the relevant definitions and equations from WDMI such that this paper is self-contained.
 
\emph{Conventions.} We adopt the cyclic frequency $f$, discretized to index $k$, as the frequency-domain argument and $t$, discretized to $j$, for the time domain.
WDM quantities are indexed by frequency $m$ and time $n$.
The flattened pair used for matrices and vectors is $(mn)$. 
Vectors and matrices are expressed in boldface.
Ensemble averages are denoted by $\langle\cdot\rangle$.
We adopt the Fourier convention of Eq.~(4) of WDMI,
\begin{equation}
    \tilde{x}(f)=\int_{-\infty}^{\infty}x(t)e^{-2\pi ift}\diff t\,,
    \qquad
    x(t)=\int_{-\infty}^{\infty}\tilde{x}(f)e^{2\pi ift}\diff f\,.
    \label{eq:fourier}
\end{equation}
The discrete DC and Nyquist Fourier bins adopt the endpoint conventions of WDMI.
Time-dependent spectra are written with time as the first argument, $S(t,f)$.
The one-sided noise PSD of WDMI is related to the two-sided spectrum $S_2(t,f)$ defined for both positive and negative frequencies via
\begin{equation}
    S(t,f)=2S_2(t,f)\,,\quad f>0\,.
    \label{eq:two-sided}
\end{equation}

\emph{Data and the time-frequency grid.} In the time-domain, the data are a real time series of duration $T_{\rm obs}$, sampled at cadence $\Delta t$ for a total of $N=T_{\rm obs}/\Delta t$ samples.
The frequency-domain data representation consists of $(N/2-1)$ complex samples for the interior frequency bins and $2$ real samples for the DC and Nyquist bins; the total remains at $2(N/2-1)+2=N$.
The time-frequency plane is tiled into $N_f$ frequency bands and $N_t$ time bins of width $\Delta F$ and $\Delta T$ respectively, satisfying
\begin{equation}
    N=N_tN_f\,,\qquad
    \Delta T=\frac{T_{\mathrm{obs}}}{N_t}\,,\qquad
    \Delta F=\frac{N}{2 T_{\mathrm{obs}} N_f}\,,\qquad
    \Delta T\,\Delta F=\frac{1}{2}\,.
    \label{eq:grid}
\end{equation}
The time-frequency cells are centered at $t_n=n\Delta T$ and $f_m=m\Delta F$ with $0\le m\le N_f$ and $0\le n\le N_t-1$.
The WDM-domain data are represented by $(N_f-1)N_t$ real samples for the interior bands, and $N_t/2$ real samples for each of the DC and Nyquist bands that are half-empty; the total again remains at $(N_f-1)N_t+2N_t/2=N$.
Since the representation is real, critical sampling and orthogonality mean that each time-frequency cell has an area of $1/2$.

\emph{The Meyer window.}
The WDM basis relies on a window that is a scaled version of the Meyer scaling function, $\tilde\phi(f)=\tilde\varphi(f)/\sqrt{\Delta F}$.
Its shape is controlled by three parameters $(d,\alpha,\beta)$, subject to $2\alpha+\beta=\Delta F$ that, together with the symmetry of the Meyer rolloff, enforces the partition-of-unity property, 
\begin{equation}
    \sum_{p\in\mathbb{Z}}\tilde\phi(f+p\Delta F)^2=\frac{1}{\Delta F}
    \qquad\Longrightarrow\qquad
    \int_{-\infty}^{\infty}\tilde\phi(f)^2\diff f=1\,.
    \label{eq:pou}
\end{equation}
The Meyer window is real and even.
Unless otherwise stated we adopt the WDMI parameters $d=4$, $\alpha=\Delta F/4$, and $\beta=\Delta F/2$, for which $\tilde\phi$ has support over one central band and $25\%$ of its neighboring bands, i.e., $f/\Delta F\in[-3/4,3/4]$.

\emph{The WDM basis and data coefficients.}
The defining property of the WDM basis is an alternating sine/cosine structure in the time-domain, Eq.~(21) of WDMI.
In the frequency-domain, the basis covers symmetrically the $\pm m\Delta F$ bands per Eq.~(16) of WDMI,
\begin{equation}
    \tilde{g}_{mn}(f)=A_me^{-2\pi inf\Delta T}\left[C_{mn}^*\tilde\phi(f+m\Delta F)+C_{mn}\tilde\phi(f-m\Delta F)\right]\,,
    \label{eq:basis}
\end{equation}
with
\begin{equation}
    A_m=\begin{cases}1/2 & m=0,N_f\\[2pt] 1/\sqrt{2} & \text{otherwise}\end{cases}\,,
    \qquad
    C_{mn}=\begin{cases}1 & m+n\ \text{even}\\[2pt] i & m+n\ \text{odd}\end{cases}\,.
    \label{eq:AmCmn}
\end{equation}
The basis is orthonormal, except for $m=0$ with $n$ odd and for $m=N_f$ with $N_f+n$ odd, where the basis elements vanish.
The data coefficients in the WDM domain are obtained by projecting the data onto the basis, Eq.~(28) of WDMI,
\begin{equation}
    w_{mn}=\int x(t)g_{mn}(t)\diff t
     =2 A_m (-1)^{mn} \Re\left[ C_{mn}^* y_{mn} \right]\,, \qquad y_{mn} = \int_{-\infty}^{\infty} \tilde{x}(f+f_m)\tilde{\phi}(f)e^{2\pi i f t_n}\diff f\,,
    \label{eq:wdm_coeffs}
\end{equation}
with $f_m = m\Delta F$ and $t_n = n\Delta T$.
Equation~\eqref{eq:wdm_coeffs} can be evaluated with $\mathcal{O}(N\log N_t)$ operations via a fast Fourier transform~\cite{Necula_2012}.

\emph{Likelihood and noise covariance.}
Packaging the WDM coefficients into a vector with flattened index $(mn)$, the transform from the time- or frequency-domains is $\mathbf{w}=\mathbf{W}\mathbf{x}=\tilde{\mathbf{W}}\tilde{\mathbf{x}}$, where the transform matrices $\mathbf{W}$ are defined through the unitary discrete Fourier transform matrix $\mathbf{F}$, $\tilde{\mathbf{W}}=\mathbf{W}\mathbf{F}^\dagger$.
Under the conventions of WDMI, both matrices are square and unitary~\cite{Pearson:2025wfd}.
However, in this work, we use the non-unitary convention and make the normalization factor explicit, $\mathbf{W}\mathbf{W}^T=\tilde{\mathbf{W}}\tilde{\mathbf{W}}^\dagger=\Delta t \,\mathbf{I}$.
The Gaussian log-likelihood in the WDM domain for residual $\mathbf{r}_{\rm w}=\mathbf{W}(\mathbf{d}-\mathbf{h}(\bm{\theta}))$ is Eq.~(110) in WDMI in the convention used here: 
\begin{equation}
    \ln p(\mathbf{d}\mid\bm{\theta})=-\frac{N}{2}\ln(2\pi) + \frac{N}{2}\ln \Delta t-\frac{1}{2}\ln\det\bm{\Lambda}
    -\frac{1}{2}\mathbf{r}_{\rm w}^T\bm{\Lambda}^{-1}\mathbf{r}_{\rm w}\,,
    \label{eq:wdm-gaussian}
\end{equation}
where the WDM noise covariance matrix is related to the time- and frequency-domain ones respectively by Eqs.~(99) and (100) in WDMI,
\begin{equation}
    \bm{\Lambda}=\mathbf{W}\mathbf{C}\mathbf{W}^T=\tilde{\mathbf{W}}\bm{\Sigma}\tilde{\mathbf{W}}^\dagger\,,
    \qquad
    \bm{\Lambda}^{-1}=\frac{1}{\Delta t^2}\tilde{\mathbf{W}}\bm{\Sigma}^{-1}\tilde{\mathbf{W}}^\dagger\,,
    \label{eq:cov-transform}
\end{equation}
where $\mathbf{C} = \langle \mathbf{n}(t)\mathbf{n}^T(t)\rangle$ is the time-domain and $\mathbf{\Sigma} = \langle \tilde{\mathbf{n}}(f)\tilde{\mathbf{n}}^*(f)\rangle$ is the frequency-domain covariance matrix.
These relations make no assumption about the noise properties.

\emph{Stationary noise and covariance structure.} For stationary noise, the WDM covariance $\bm{\Lambda}$ is block tri-diagonal.
If, in addition, the spectrum is flat across the three bands spanned by the Meyer filter, the covariance is diagonal, $\Lambda_{(mn)(m'n')}\approx\tfrac{1}{2}S(f_m)\delta_{mm'}\delta_{nn'}$, Eq.~(70) in WDMI.
If the spectrum varies slowly in time, on the scale of $\Delta T$, the WDM covariance is Eq.~(118) in WDMI,
\begin{equation}
    \Lambda_{(mn)(m'n')}\approx\tfrac{1}{2}S(t_n,f_m)\,\delta_{mm'}\delta_{nn'}
    \equiv\tfrac{1}{2}S_{mn}\,\delta_{mm'}\delta_{nn'}\,,
    \label{eq:midpoint}
\end{equation}
evaluated at the WDM cell midpoints.
WDMI introduced Eq.~\eqref{eq:midpoint} as a working definition of the evolutionary spectrum.
Its precise properties are the focus of this paper.

\section{Introduction to Evolutionary Spectra}
\label{sec:evolutionary}

The theory of evolutionary spectra is summarized from \citet{Priestley1981}.\footnote{That literature uses angular, instead of cyclic, frequency. Our two-sided spectrum is related to the relevant spectrum in \citet{Priestley1981} with $S_2(t,f)=2\pi H(t,2\pi f)$. Throughout this section, we adopt the WDM notation.}
Our goal in this section is to provide a practical discussion of nonstationary processes, and defer to the original text for formal derivations.
The spectrum of a nonstationary process needs to satisfy certain conditions: (i) it is a time-dependent function, (ii) it can be computed from a time-series realization of the process alone,\footnote{While most analyses model the noise spectrum during inference rather than compute it from the data, this requirement remains: it must be possible, at least in theory, to compute the spectrum from a realization with no additional knowledge about the underlying noise properties.} (iii) it reduces to the usual spectrum for stationary processes, and (iv) it has a similar physical interpretation to the stationary-process spectrum, namely it is a local distribution of power over frequency at each time.
Evolutionary spectra satisfy these conditions.

\subsection{Intuition and definition of evolutionary spectra}
\label{sec:evolutionary-definition}

The requirement that the process spectrum can be interpreted as a power-frequency distribution is nontrivial and guides the definition of the evolutionary spectrum.
To understand the complications arising from nonstationarity, we start with a (continuous) stationary process $x(t)$, in $-\infty<t<\infty$ with a vanishing mean and finite variance, $\left\langle x(t)\right\rangle = 0,\,
\left\langle |x(t)|^2\right\rangle < \infty$.
The spectrum interpretation follows from the fact that the process can be expressed in spectral form, i.e., a sum of sines and cosines with random amplitudes and phases~\cite{Cramer1951}:
\begin{equation}
    x(t)=\int_{-\infty}^{+\infty} e^{2\pi i f t} \tilde{x}(f) \diff f\,,
    \label{eq:stationary-def}
\end{equation}
where $\diff \tilde{x}(f)\equiv \tilde{x}(f) \diff f$ represents the random complex amplitude (magnitude and phase) of frequencies $[f,f+\diff f]$.
For real data, $\diff \tilde{x}(-f) =\diff \tilde{x}^*(f)$.
For a stationary process, $\diff \tilde{x}(f)$ are orthogonal across frequency bands, $\left\langle \diff \tilde{x}(f)\diff \tilde{x}^*(f') \right\rangle=S_2(f) \delta(f-f') \diff f \diff f'$.
Under this expression, $f$ is interpreted as the ``frequency'' of an ``oscillatory'' feature, and since the $\diff \tilde{x}(f)$ are orthogonal, we can unambiguously identify the component of $x(t)$ that oscillates at each band and define its power.
The covariance kernel is
\begin{equation}
R(t, t') = \left\langle x(t) x^*(t')\right\rangle\,,
\end{equation}
and substituting Eq.~\eqref{eq:stationary-def} we recover the Wiener-Khinchin theorem~\cite{Wiener1930,Khinchin1934}:
\begin{align}
R(t, t') &= \left\langle \int_{-\infty}^{+\infty} e^{2\pi i f t}\tilde{x}(f) \diff f \int_{-\infty}^{+\infty} e^{-2\pi i f' t'}\tilde{x}^*(f') \diff f'\right\rangle\nonumber\\
&=\int_{-\infty}^{+\infty}\int_{-\infty}^{+\infty} e^{2\pi i (f t-f't')}\left\langle \diff \tilde{x}(f)\diff \tilde{x}^*(f')\right\rangle\nonumber \\
&=\int_{-\infty}^{+\infty}\int_{-\infty}^{+\infty} e^{2\pi i (f t-f't')}S_2(f) \delta(f-f') \diff f \diff f'\nonumber\\
&=\int_{-\infty}^{+\infty} e^{2\pi i f (t-t')}S_2(f) \diff f\,.
\end{align}
The covariance $R(t, t')$ is a function of the time-difference $|t-t'|$ alone, with $S_2(f)\diff f$ the power in $[f,f+\diff f]$ and $S_2(f)$ the spectral density.
In other words, orthogonality of $\diff \tilde{x}(f)$ and a constant-amplitude basis of $\{e^{2\pi i f t}\}$ is equivalent to $R(t, t')=R(|t-t'|)$ and thus a stationary process.
Reversing the logic, a nonstationary process cannot be expressed in the form of Eq.~\eqref{eq:stationary-def} with $\diff \tilde{x}(f)$ uncorrelated across frequencies and it is no longer straightforward to define the $x(t)$ component at a given frequency.
Instead, a nonstationary process must either admit nonorthogonal frequency bands (a nondiagonal frequency-domain covariance matrix), or generalize the basis. 
Evolutionary spectra take the second route. 

The idea behind evolutionary spectra is to generalize the notion of ``frequency'' of an ``oscillatory'' feature beyond constant-amplitude sines and cosines.
Priestley offers the following illustrative (though deterministic) example: 
\begin{equation}
x(t) = A \, e^{-t^2/\tau^2} \cos(2\pi f_0 t + \theta_0)\,.
\end{equation}
Under a conventional sine/cosine Fourier transform, $x(t)$ contains components at all frequencies, with power peaking at $\pm f_0$ and width inversely proportional to $\tau$.
An alternative interpretation is that $x(t)$ contains components only at $\pm f_0$ but with a decaying amplitude of $A \, e^{-t^2/\tau^2}$.
The more intuitive ``oscillatory'' functions on which to define frequencies in this case are the Gaussian-decaying sine/cosines.
Of course not every function can be interpreted as ``oscillatory,'' heuristically the requirement being that its Fourier transform peaks at a single frequency and has a finite width.
In that case, it behaves locally like a sinusoid, but with a smoothly-varying amplitude.

Generalizing the sines/cosines of $\{e^{2\pi i f t}\}$ to oscillatory functions with time-dependent amplitudes, $\{\psi(t,f)\}$, the nonstationary process is
\begin{equation}
    x(t)=\int_{-\infty}^{+\infty} \psi(t,f) \hat{x}(f)\diff f\,,
    \label{eq:nonstationary-def}
\end{equation}
where $\diff \hat{x}(f)$ are again orthogonal with $\left\langle \diff \hat{x}(f)\diff \hat{x}^*(f') \right\rangle={S_{2,\psi}}(f) \delta(f-f')\diff f \diff f'$ and covariance kernel
\begin{align}
R(t, t') 
&=\int_{-\infty}^{+\infty} {\psi(t,f)} {\psi^*(t',f)} {S_{2,\psi}}(f)\diff f \,,
\end{align}
where now ${S_{2,\psi}}(f)$ plays the role of the spectral density, and the $\psi$ subscript denotes the oscillatory family.
Both $\hat{x}$ and $S_{2,\psi}$ depend on the choice of $\{\psi(t,f)\}$ and are distinct from their Fourier-transform counterparts.
The choice of $\{{\psi(t,f)}\}$ is not unique.
For stationary processes, ${\psi(t,f)}=e^{2\pi i f t}$ is a natural choice and provides the common physical interpretation of the spectrum as a power-frequency distribution.
For nonstationary processes that can be expressed as modulated sines/cosines,
\begin{equation}
{\psi(t,f)}=A(t,f)\,e^{2\pi i f t}\,,
\label{eq:oscillatory_family}
\end{equation}
where $A(t,f)$ is a slowly-varying envelope, i.e., its Fourier transform with respect to $t$ is concentrated at zero frequency,
the covariance becomes 
\begin{align}
R(t, t') 
&=\int_{-\infty}^{+\infty} A(t,f) A^*(t',f)\,e^{2\pi i f (t-t')} {S_{2,\psi}}(f) \diff f \,.
\label{eq:evolutionary-covariance}
\end{align}
A process that admits such a representation is labeled an ``oscillatory'' process and its total power at time $t$ is
\begin{equation}
\left\langle |x(t)|^2\right\rangle=R(t, t)=\int_{-\infty}^{+\infty} |A(t,f)|^2  {S_{2,\psi}}(f) \diff f\,,
\end{equation}
from which we can immediately read off the (nonnegative by construction) distribution of power over frequency at a time instant, namely the \emph{evolutionary power spectrum}~\cite{Priestley1965Evolutionary}
\begin{equation}
 S_2(t,f) \diff f=|A(t,f)|^2  {S_{2,\psi}}(f) \diff f\,,
 \label{eq:evolutionary_envelope_magnitude}
\end{equation}
and the \emph{evolutionary spectral density} $S_2(t,f)$.
Even though the evolutionary power spectrum in Eq.~\eqref{eq:evolutionary_envelope_magnitude} is tied to a specific choice of $\{\psi(t,f)\}$, its integral represents the total power and thus it is independent of that choice.
It also immediately satisfies three of the required conditions: it is time-dependent, it is by construction a local distribution of power over frequency, and it reduces to the normal power spectrum for stationary processes if ${\psi(t,f)}=e^{2\pi i f t}$.
The remaining condition, namely that the spectrum can be computed from a single realization of the process, and the awkward dependence on the choice of ${\psi(t,f)}$ are the subject of Sec.~\ref{sec:evolutionary-estimation}.

\subsection{Evaluating the evolutionary spectrum}
\label{sec:evolutionary-estimation}

The meaningful definition and estimation of the evolutionary spectrum relies on $A(t,f)$ being a slowly-varying amplitude, namely the process is \emph{semi-stationary}.
In that case, the spectrum can be estimated by standard filtering and any choice of $\{{\psi(t,f)}\}$ yields the same measurable spectrum, up to corrections of order the ratio of the filter and the nonstationarity timescales.
The problem contains the following relevant scales:
\begin{enumerate}
\item $T_{\rm obs}$, the length of the process realization from which we are estimating the evolutionary power spectrum;
\item $B_x$, the largest $A(t,f)$ variation timescale over admissible functions $\{{\psi(t,f)}\}$, effectively the width of the semi-stationary process interpreted as the maximum time interval over which the process is approximately stationary;
\item $B_{\phi}$, the temporal extent of the filter $\phi$ used for spectral estimation.
\end{enumerate}
See \citet{Priestley1981} for the formal definition of the scales in terms of first moments.
The following construction and operational definition of the evolutionary power spectrum demand $B_{\phi}\ll B_x \ll T_{\rm obs}$: we have a long stretch of data whose spectrum varies slowly with respect to the filter used for estimation.

Given a process $x(t)$, the power near a frequency $f_0$ can be estimated by band-passing,
\begin{equation}
    y(t;f_0)=\int_{-\infty}^{+\infty} \phi(u)\, x(t-u)\, e^{-2\pi i f_0 (t-u)}\,\diff u\,,
    \label{eq:demodulate}
\end{equation}
implemented here as a complex demodulation that first band-shifts from $f_0$ to the DC band and then applies a low-pass filter $\phi(u)$ with transfer function
\begin{equation}
    \tilde{\phi}(f)=\int_{-\infty}^{+\infty} \phi(u)\, e^{-2\pi i f u}\,\diff u\,,
    \label{eq:transfer}
\end{equation}
concentrated around the DC band and normalized as $\int|\tilde{\phi}(f)|^2\diff f=\int|\phi(u)|^2\diff u=1$.\footnote{The filter is scaled by $\sqrt{2\pi}$ relative to Priestley's angular-frequency normalization.}
For a stationary process, the spectrum of the demodulated process is related to the underlying process spectrum as
\begin{equation}
S_2^{(y)}(f;f_0)\diff f=\left|\tilde{\phi}(f)\right|^2S_2^{(x)}(f+f_0)\diff f\,.
    \label{eq:demodulated_spectrum}
\end{equation}
Thus applying a filter on the data with known transfer function $\tilde{\phi}(f)$ allows us to estimate the power spectrum of the original process at different values of $f_0$.

For a nonstationary process that satisfies the timescale hierarchy, the process envelope $A(t,f)$ is approximately constant across the filter for any admissible choice of $\{\psi(t,f)\}$ and for all $f$.
The band-pass power can then be shown to be \cite{Priestley1965Evolutionary}
\begin{align}
\left\langle |y(t;f_0)|^2\right\rangle&=\int_{-\infty}^{+\infty}\left|\tilde{\phi}(f)\right|^2S_2^{(x)}(t,f+f_0)\diff f + {\cal{O}}\left(B_{\phi}/B_x\right)\,,
\label{eq:filter-power}\\
&\simeq\int_{-\infty}^{+\infty}\left|\tilde{\phi}(f-f_0)\right|^2S_2^{(x)}(t,f)\diff f\,.
    \label{eq:filter-power-convolution}
\end{align}
In the second line, we have used a change of variable and the fact that $\tilde{\phi}$ is even to write the power as a convolution between the process spectrum and the filter (squared) evaluated at $f_0$.
The band-passed filter returns the power of the original process at different values of $f_0$ \emph{averaged over the filter band $\sim1/B_\phi$} (as we only have access to the integral of Eq.~\eqref{eq:filter-power} and not the integrand) and at the present time $t$.
Sweeping across $f_0$, we obtain the full band-averaged spectrum, which is the measurable quantity (not $S_2^{(x)}(t,f)$ itself).

Besides offering a way to estimate the evolutionary power spectrum through standard filtering, Eq.~\eqref{eq:filter-power} also reveals that to leading order in $B_{\phi}/B_x$, the band-average power is \emph{independent of the choice of oscillatory functions $\{{\psi(t,f)}\}$}.
The left-hand side of Eq.~\eqref{eq:filter-power} is directly evaluated from the process realization without specifying a family $\{{\psi(t,f)}\}$.
The right-hand side then suggests that any choice of $\{{\psi(t,f)}\}$ whose envelope varies on timescale longer than $B_{\phi}$ would give the same band-average power for scales larger than what the filter can resolve, namely $B_{\phi}$ in time and $1/B_{\phi}$ in frequency.
These considerations also reveal an uncertainty principle limit: a short-in-time filter loses spectral resolution as it averages over a wider band.

In practice, we have a finite-duration realization of the process $T_{\rm obs}$.
If the process was stationary, we would be able to extend the filter over the entire duration $T_{\rm obs}$ and improve spectral resolution up to $1/T_{\rm obs}$.
But for a nonstationary process the temporal extent of the filter is restricted by $B_{\phi}\ll B_x$.
The power $|y(t;f_0)|^2$ will then be an unbiased estimator of the band-averaged power of Eq.~\eqref{eq:filter-power} with a variance for Gaussian $x(t)$ of
\begin{equation}
\mathrm{Var}\big[|y(t;f_0)|^2\big]\simeq\left(\left\langle |y(t;f_0)|^2\right\rangle\right)^2\,,
\end{equation}
which is 100\% variance independent of how much data $T_{\rm obs}$ we have.
To improve the accuracy of the power estimate and reduce fluctuations, we average the power around $t$ for $Q$ filters and duration $T_q\simeq Q B_\phi$, larger than the filter extent but smaller than the nonstationarity timescale, with window $q(u)$ normalized as $\int q(u)\diff u = 1$.
Thus sacrificing some time resolution, we obtain
\begin{equation}
    P_q(t;f_0)=\int_{-\infty}^{+\infty} q(u)\, |y(t-u;f_0)|^2\, \diff u\,.
    \label{eq:time-smoothing}
\end{equation}
The mean of $P_q(t;f_0)$ is the band-averaged spectral density of Eq.~\eqref{eq:filter-power} averaged over $T_q$ but its variance has been suppressed to~\cite{Priestley1966}
\begin{equation}
    \mathrm{Var}\left[P_q(t;f_0)\right] \simeq \frac{C}{2\pi T_q}\, \left(\left\langle P_q(t;f_0)\right\rangle\right)^2 \int_{-\infty}^{+\infty}|\tilde{\phi}(f)|^4\, \diff f\,,
    \label{eq:smoothed-variance}
\end{equation}
where $C$ is an ${\cal{O}}(1)$ constant that depends on the smoothing filter.
Since the transfer function integral scales with the filter length, the variance scales inversely with the number of filters averaged over.

Overall, the evolutionary spectrum can be estimated from a process realization satisfying the final condition (ii), with a relevant timescale hierarchy of $B_{\phi}\ll T_q \ll B_x \ll T_{\rm obs}$:
we have enough data to probe the nonstationarity and the filter is short enough compared to the nonstationarity that the spectrum does not vary across many filter lengths used for averaging to suppress sample variance.
In the $B_x\to \infty$ limit of a stationary process, we recover the standard Welch estimate.

\section{Evolutionary spectra in the WDM basis}
\label{sec:evolutionary-wdm}

In this section, we apply the framework of evolutionary power spectra of Sec.~\ref{sec:evolutionary} to the WDM basis.
The main product is the WDM noise covariance matrix given an evolutionary spectrum, $S(t,f)$, under the timescale hierarchy of Sec.~\ref{sec:evolutionary}: Eq.~\eqref{eq:wdm_variance_band_average} for the diagonal and Eq.~\eqref{eq:offdiag_covariance} for the off-diagonal parts.
These results are subject to two interpretations depending on the application.
The first application is the one discussed in Sec.~\ref{sec:evolutionary-estimation}: given some realization for the process, we want to estimate the evolutionary spectral density. 
In this case, we calculate the left-hand side of Eq.~\eqref{eq:wdm_variance_band_average}, $\langle w_{mn}^2\rangle$, from the realization and the right-hand side allows us to interpret the result as the band-averaged evolutionary spectrum.
The second application is inference: the evolutionary spectral density is modeled.
In that case, given a model for $S(t,f)$, Eqs.~\eqref{eq:wdm_variance_band_average} and~\eqref{eq:offdiag_covariance} give the noise covariance $\Lambda_{(mn)(m'n')}=\langle w_{mn}w_{m'n'}\rangle$ which is then used to compute the likelihood.

In the framework of Sec.~\ref{sec:evolutionary}, the WDM basis amounts to a critically-sampled collection of Meyer filters, each centered at the demodulation frequency $f_0=f_m=m\Delta F$.
The complex demodulate of Eq.~\eqref{eq:demodulate} evaluated on the WDM grid $(t=t_n,f_0=f_m)$ becomes
\begin{equation}
    y_{mn} = \int_{-\infty}^{\infty} \phi(u)\,x(t_n-u)\,e^{-2\pi i f_m(t_n-u)}\diff u\,,
    \label{eq:wdm_complex_demodulate}
\end{equation}
where the low-pass filter is the (real) Meyer window that automatically satisfies $\int{\tilde{\phi}(f)^2}\diff f=1$, Eq.~\eqref{eq:pou}.
This is identical to the $y_{mn}$ factor that enters the WDM coefficients in Eq.~\eqref{eq:wdm_coeffs}, as can be shown by a short calculation of change of variables after substituting the $x(t)$ Fourier transform.
The WDM coefficients of Eq.~\eqref{eq:wdm_coeffs} are therefore the real or imaginary part of the demodulate $y_{mn}$, determined by $C^*_{mn}$ and the parity of $m+n$. 

\subsection{Diagonal covariance terms}
\label{sec:diagonal}

The variance of the WDM coefficients is 
\begin{align}
    \langle w_{mn}^2\rangle&=\left\langle \left(2A_m(-1)^{mn}\Re[C_{mn}^*y_{mn}]\right)^2\right\rangle\nonumber\\
    &=2A_m^2\left(\langle |y_{mn}|^2\rangle+\Re\left[(C_{mn}^*)^2\,\langle y_{mn}^2\rangle\right]\right)\,,
    \label{eq:wdm_demodulate_second_moment}
\end{align}
where we have pulled deterministic terms out of the expectation value, and used $|C_{mn}|=1$, $(-1)^{2mn}=1$, and the complex identity $(\Re z)^2=(|z|^2+\Re z^2)/2$.
The two terms of Eq.~\eqref{eq:wdm_demodulate_second_moment} are the expectation value of the demodulate squared, with and without a complex conjugation. 
Defining $y^{(-1)}_{mn}\equiv y^*_{mn}$ and $y^{(+1)}_{mn}\equiv y_{mn}$ such that $\sigma=\pm 1$ controls the conjugation, both terms can be expressed as
\begin{align}
    \left\langle y_{mn}y^{(\sigma)}_{mn}\right\rangle
    &=\iint\phi(s)\phi(s')\,\langle x(t_n-s)x(t_n-s')\rangle\,e^{2\pi i f_m(s+\sigma s')}\diff s\diff s'\nonumber\\
    &=\iint\phi(s)\phi(s')\,R(t_n-s,t_n-s')\,e^{2\pi i f_m(s+\sigma s')}\diff s\diff s'\,.
    \label{eq:wdm_moment_exact}
\end{align}
Substituting the covariance for an evolutionary spectrum, Eq.~\eqref{eq:evolutionary-covariance}, we obtain
\begin{align}
    \left\langle y_{mn}y^{(\sigma)}_{mn}\right\rangle=\int S_{2,\psi}(f)
    &\left[\int\phi(s)A(t_n-s,f)e^{-2\pi i(f-f_m)s}\diff s\right]\left[\int\phi(s')A^*(t_n-s',f)e^{2\pi i(f+\sigma f_m)s'}\diff s'\right]\diff f\,.
    \label{eq:wdm_exact_response}
\end{align}
To evaluate the two interior integrals, we use the timescale approximation of Sec.~\ref{sec:evolutionary-estimation}.
The noise varies slowly compared to the extent of the Meyer filter, $B_\phi\ll B_x$, and thus $A(t_n-s,f)\simeq A(t_n,f)$ and $A(t_n-s',f)\simeq A(t_n,f)$.
The two integrals, therefore, amount to a Fourier transform of the filter,
\begin{equation}
    \left\langle y_{mn}y^{(\sigma)}_{mn}\right\rangle
    \simeq\int_{-\infty}^{\infty} S_2(t_n,f)\,\tilde{\phi}(f-f_m)\,\tilde{\phi}(f+\sigma f_m)\diff f\,,
    \label{eq:wdm_moment_frozen}
\end{equation}
where we have used the fact that the filter is even and substituted $S_2(t_n,f)=|A(t_n,f)|^2S_{2,\psi}(f)$ from Eq.~\eqref{eq:evolutionary_envelope_magnitude}.

At this point, we need to separate the two $\sigma$ cases.
For $\sigma=-1$, we obtain the power of the demodulate, band-averaged by the Meyer filter,
\begin{equation}
    \left\langle \left|y_{mn}\right|^2\right\rangle
    \simeq\int_{-\infty}^{\infty} S_2(t_n,f)\,\tilde{\phi}(f-f_m)^2\diff f\,.
    \label{eq:wdm_band-average}
\end{equation}
For $\sigma=+1$, $\tilde{\phi}(f-f_m)$ and $\tilde{\phi}(f+f_m)$ do not overlap for the interior bands, $m\neq0,N_f$. 
At the DC band, $m=0$, the integral is identical to its $\sigma=-1$ counterpart.
At Nyquist, $m=N_f$, the two terms again coincide due to the $2f_\mathrm{Nyq}$ periodicity of the discrete data shown in Appendix~C of WDMI.
Hence
\begin{equation}
    \langle y^2_{mn}\rangle\simeq
    \begin{cases}
        0\,, & 1\le m\le N_f-1\,,\\[2pt]
        \langle |y_{mn}|^2\rangle\,, & m=0,N_f\,.
    \end{cases}
    \label{eq:wdm_unconjugated_moment}
\end{equation}

Substituting Eqs.~\eqref{eq:wdm_unconjugated_moment} and~\eqref{eq:wdm_band-average} into Eq.~\eqref{eq:wdm_demodulate_second_moment}, with $2A_m^2=1$ for the interior bands and $2A_m^2=1/2$ for the edge ones, and setting $(C_{mn}^*)^2=(-1)^{m+n}$, we obtain the WDM variance
\begin{equation}
    \langle w_{mn}^2\rangle
    \simeq{\varepsilon_{mn}}\int_{-\infty}^{\infty}
       S_2(t_n,f){\tilde{\phi}(f-f_m)^2}\diff f=\frac{{\varepsilon_{mn}}}{2}\int_{-\infty}^{\infty}
       S(t_n,f){\tilde{\phi}(f-f_m)^2}\diff f\,,
    \label{eq:wdm_variance_band_average}
\end{equation}
with parity
\begin{equation}
    \varepsilon_{mn}=
    \begin{cases}
        1\,, & 1\le m\le N_f-1\,,\\[2pt]
        \tfrac{1}{2}\left[1+(-1)^{m+n}\right]\,, & m=0,N_f\,.
    \end{cases}
    \label{eq:wdm_edge_weight}
\end{equation}
For the edge bands $\varepsilon_{mn}=1$ for $m+n$ even and $\varepsilon_{mn}=0$ for $m+n$ odd; the half-empty DC and Nyquist bands of Sec.~\ref{sec:notation} are thus recovered.
Up to the parity correction, Eq.~\eqref{eq:wdm_variance_band_average} is identical to the evolutionary band-passed power of Eq.~\eqref{eq:filter-power}, $\langle w_{mn}^2\rangle=\langle |y_{mn}|^2\rangle$, evaluated on the WDM grid of $t=t_n$ and $f_0=f_m$.
Partition-of-unity, Eq.~\eqref{eq:pou}, also guarantees that the band-averaged spectrum conserves the total power at each time slice $t_n$,
\begin{equation}
    2 \Delta F\sum_{m}\left\langle w_{mn}^2\right\rangle=\int_{-\infty}^{\infty}S_2(t_n,f)\diff f\,.
    \label{eq:power_conservation}
\end{equation}
If the spectrum is also approximately constant across the frequency support of the window, the WDM variance reduces to~\cite{Cornish:2020odn,Vajpeyi:2026msr,Johnson:2026rrn}
\begin{equation}
    \langle w_{mn}^2\rangle\simeq {\varepsilon_{mn}}S_2(t_n,f_m) = \frac{{\varepsilon_{mn}}}{2}S(t_n,f_m)\,.
    \label{eq:wdm_variance_cell_center}
\end{equation}
Equation~\eqref{eq:wdm_variance_band_average} is the leading-order diagonal term of the WDM noise covariance matrix to ${\cal{O}}(B_\phi/B_x)$ and we refer to it as the ``band-averaged'' covariance.
Equation~\eqref{eq:wdm_variance_cell_center} is instead referred to as the ``midpoint'' covariance, because it is evaluated at the center of the time-frequency cells.

\subsection{Off-diagonal covariance terms}
\label{sec:offdiagonal}

The demonstration of Sec.~\ref{sec:evolutionary-estimation} that the band-average power is independent of the choice of oscillatory functions to represent the nonstationary process concerns $\left\langle w_{mn}^2\right\rangle$, i.e., the covariance matrix diagonal terms.
Retracing the steps of Sec.~\ref{sec:diagonal}, the dependence on the oscillatory functions disappears from Eq.~\eqref{eq:wdm_exact_response} to Eq.~\eqref{eq:wdm_moment_frozen}.
Specifically, fixing the oscillatory function envelopes across the filter identifies the term $A(t_n,f)A^*(t_n,f)S_{2,\psi}(f)=|A(t_n,f)|^2S_{2,\psi}(f)$ which is the evolutionary spectral density $S_2(t_n,f)$ from Eq.~\eqref{eq:evolutionary_envelope_magnitude}. 

A similar calculation for the off-diagonal terms, $\langle w_{mn}w_{m'n'}\rangle$, yields after decomposing the envelope into an argument and a phase, $A(t,f)=\left|A(t,f)\right|e^{i\theta(t,f)}$,
\begin{align}
    \left\langle y_{mn}y^{(\sigma)}_{m'n'}\right\rangle
    &\simeq\int S_{2,\psi}(f)e^{2\pi i\left[(f-f_m)t_n-(f+\sigma f_{m'})t_{n'}\right]}
    A(t_n,f)A^*(t_{n'},f)\tilde\phi(f-f_m)\tilde\phi(f+\sigma f_{m'})\diff f\nonumber \\
    &=\int S_{2,\psi}(f)e^{2\pi i\left[(f-f_m)t_n-(f+\sigma f_{m'})t_{n'}\right]}
    \left|A(t_n,f)\right|\left|A(t_{n'},f)\right|e^{i\left[\theta(t_n,f)-\theta(t_{n'},f)\right]}\tilde\phi(f-f_m)\tilde\phi(f+\sigma f_{m'})\diff f\,.
    \label{eq:wdm_expectation_offdiagonal}
\end{align}  
The envelope amplitude is absorbed in the evolutionary spectrum but the phase difference is not.
A residual phase dependence would mean that for the same process, two oscillatory families will result in the same evolutionary spectra and diagonal terms but different off-diagonal terms.

However, the timescale expansion of Sec.~\ref{sec:evolutionary-estimation} shows that the phase dependence is also ${\cal{O}}(B_\phi/B_x)$ and thus of the same order as terms already neglected from the diagonals.
The covariance between $t_n$ and $t_{n'}$ is non-negligible only if the two corresponding filters overlap in time, $|t_n-t_{n'}|\lesssim B_\phi$, while the envelope amplitude and phase vary on the nonstationarity timescale $B_x$. 
Expanding the amplitude and phase terms of Eq.~\eqref{eq:wdm_expectation_offdiagonal} about the time midpoint $\bar{t}=(t_n+t_{n'})/2$ with $\delta t=(t_n-t_{n'})/2$, we obtain
\begin{align}
    \left|A(\bar{t}+\delta t,f)\right|\left|A(\bar{t}-\delta t,f)\right|
    &=\left|A(\bar{t},f)\right|^2\left[1+{\cal{O}}\left(\frac{\delta t^2}{B_x^2}\right)\right]\,,\nonumber\\
    \theta(\bar{t}+\delta t,f)-\theta(\bar{t}-\delta t,f)
    &=2\delta t\,\partial_t\theta(\bar{t},f)+{\cal{O}}\left(\delta t^3\right)\lesssim\frac{B_\phi}{B_x}\,.
    \label{eq:offdiag_expansion}
\end{align}
In the first equation we have set $\partial_t A/A \simeq 1/B_x$ and in the second equation we have used $\delta t\lesssim B_\phi$ and again $\partial_t\theta\lesssim1/B_x$. 
The full envelope term of Eq.~\eqref{eq:wdm_expectation_offdiagonal} is
\begin{equation}
    A(t_n,f)A^*(t_{n'},f)=\left|A(\bar{t},f)\right|^2\left[1+{\cal{O}}\left(\frac{B_\phi}{B_x}\right)\right]\,,
    \label{eq:offdiag_family_independence}
\end{equation}
and thus the off-diagonal terms are family-independent to the same order as the diagonal ones.
Following similar steps as Eq.~\eqref{eq:wdm_moment_frozen}, Eq.~\eqref{eq:wdm_expectation_offdiagonal} becomes
\begin{align}
    \left\langle y_{mn}y^{(\sigma)}_{m'n'}\right\rangle
    &\simeq \int S_2(\bar{t},f)e^{2\pi i\left[(f-f_m)t_n-(f+\sigma f_{m'})t_{n'}\right]}
    \tilde\phi(f-f_m)\tilde\phi(f+\sigma f_{m'})\diff f\nonumber\\
    &\simeq (-1)^{mn+m'n'}\int S_2(\bar{t},f)e^{2\pi if(t_n-t_{n'})}
    \tilde\phi(f-f_m)\tilde\phi(f+\sigma f_{m'})\diff f\,.
    \label{eq:offdiag_moment_frozen}
\end{align} 
Defining 
\begin{equation}
\begin{aligned}
    I^{(\sigma)}_{(mn)(m'n')}\equiv(-1)^{mn+m'n'}\left\langle y_{mn}y^{(\sigma)}_{m'n'}\right\rangle
    &=\int \bigg[ S_2(\bar{t},f)e^{2\pi if(t_n-t_{n'})}
    \tilde\phi(f-f_m)\tilde\phi(f+\sigma f_{m'})
   \bigg]  \diff f\,
    \label{eq:offdiag_I}
\end{aligned}
\end{equation}
the covariance becomes
\begin{equation}
    \Lambda_{(mn)(m'n')}\simeq2A_mA_{m'}\,
    \Re\left[C^*_{mn}C_{m'n'}I^{(-1)}_{(mn)(m'n')}
    +C^*_{mn}C^*_{m'n'}I^{(+1)}_{(mn)(m'n')}\right]\,.
    \label{eq:offdiag_covariance}
\end{equation}
which reduces to the diagonal form of Eq.~\eqref{eq:wdm_demodulate_second_moment} for $(m'n')=(mn)$.

Equation~\eqref{eq:offdiag_covariance} is the covariance for nonstationary noise to leading order in $B_\phi/B_x$ given its evolutionary power spectrum.
The $I^{(+1)}_{(mn)(m'n')}$ term contains $\tilde{\phi}(f-f_m)\tilde{\phi}(f+f_{m'})$ which is nonvanishing only if one of the bands is the DC or Nyquist bands: $(m,m')=(0,1)$ and $(N_f,N_f-1)$.
The $I^{(-1)}_{(mn)(m'n')}$ term contains $\tilde{\phi}(f-f_m)\tilde{\phi}(f-f_{m'})$, which is nonvanishing only for $|m-m'|\le1$: the covariance is block tri-diagonal at this order, as was the case for stationary noise.
The time indices $n$ and $n'$ enter through $t_n-t_{n'}$ in the exponential and through $\bar{t}=(t_n+t_{n'})/2$ in the spectrum: each $mm'$ block is approximately Toeplitz in $n-n'$, modulated by the time-dependence of $S_2(\bar{t},f)$.

\section{Generating the exact covariance}
\label{sec:cov-generation}

In Sec.~\ref{sec:evolutionary-wdm} we calculated the WDM noise covariance to leading order in the timescale hierarchy, $B_\phi/B_x$, which eliminates the oscillatory family dependence and provides ready-to-use expressions for inference and likelihood estimation.
However, this is an approximation designed to ensure a physically meaningful definition of the evolutionary power spectrum, and it does not exactly match the noise generation process.
In order to simulate data, we therefore in this section calculate the WDM covariance exactly, assuming no timescale hierarchy.
The price to pay is the need to specify an oscillatory family, but this is only required in order to simulate data. 
Analysis of data then proceeds with the results of Sec.~\ref{sec:evolutionary-wdm}, requiring modeling only $S(t,f)$.

\subsection{The exact covariance}
\label{sec:exact-covariance}

Projecting the oscillatory process of Eqs.~\eqref{eq:nonstationary-def} and~\eqref{eq:oscillatory_family} onto the WDM basis of Eq.~\eqref{eq:wdm_coeffs} yields 
\begin{equation}
    w_{mn}=\int g_{mn}(t)x(t)\diff t=\int K_{mn}(f)\diff\hat{x}(f)\,,
    \qquad
    K_{mn}(f)=\int g_{mn}(t)A(t,f)e^{2\pi ift}\diff t\,,
    \label{eq:exact_kernel}
\end{equation}
where $K_{mn}(f)$ is a deterministic kernel.
Since $\diff\hat{x}(f)$ are orthogonal and $w_{m'n'}$ are real, the covariance is
\begin{align}
    \Lambda_{(mn)(m'n')}=\left\langle w_{mn}w_{m'n'}\right\rangle
    &=\iint K_{mn}(f)K^*_{m'n'}(f')\,\left\langle \diff\hat{x}(f)\diff\hat{x}^*(f')\right\rangle\nonumber\\
    &=\int_{-\infty}^{\infty}K_{mn}(f)K^*_{m'n'}(f)\,S_{2,\psi}(f)\diff f\nonumber \\
    &=2\Re\int_{0}^{\infty}K_{mn}(f)K^*_{m'n'}(f)S_{2,\psi}(f)\diff f\nonumber \\
    &=\Re\int_{0}^{\infty}k_{mn}(f)k^*_{m'n'}(f)\diff f\,,
    \label{eq:exact_cov}
\end{align}
where in the third line we have used the fact that the process is real and thus $A(t,-f)=A^*(t,f)$ and $K_{mn}(-f)=K^*_{mn}(f)$, $g_{mn}$ is real, and $S_{2,\psi}$ is even, while in the last line we have defined
\begin{equation}
    k_{mn}(f)\equiv\sqrt{2S_{2,\psi}(f)}\,K_{mn}(f)\,.
\end{equation}
No timescale assumption has been made about the noise at this point.
Splitting again the envelope in magnitude and phase, and using the definition of the evolutionary spectrum in Eq.~\eqref{eq:evolutionary_envelope_magnitude}, we find
\begin{equation}
    k_{mn}(f)=\int_{-\infty}^{\infty}g_{mn}(t)\sqrt{S(t,f)}\,e^{i\theta(t,f)}e^{2\pi ift}\diff t\,.
    \label{eq:k_definition}
\end{equation}
This result is the exact version of the calculations of Sec.~\ref{sec:evolutionary-wdm}: the exact covariance depends on the oscillatory family through $\theta(t,f)$, but this dependence vanishes to leading order in $B_\phi/B_x$.
Equation~\eqref{eq:exact_cov} is (the real part of) a Gram matrix in $k_{mn}$ and thus the exact covariance is positive semi-definite by construction.
There is no such guarantee for the leading-order expressions of Sec.~\ref{sec:evolutionary-wdm}.

This Gramian structure also motivates the nonstationary noise model of \citet{Cornish:2026tjt}.
That model was obtained by defining a positive dynamic spectrum $S(f,t)=s^2(f,t)$, which generates a positive semi-definite covariance by construction, and yields Fourier- and time-frequency-domain covariance matrices of Gramian form.
The Gramian construction is an oscillatory process in the \citet{Priestley1981} sense, with the identification $s(f,t)\rightarrow\left|A(t,f)\right|\sqrt{2S_{2,\psi}(f)}=\sqrt{S(t,f)}$ in Eq.~\eqref{eq:k_definition} and time-independent $\theta(t,f)$.
Casting the problem in Priestley's language instead allows us to rely on the theory of Sec.~\ref{sec:evolutionary-estimation} and especially the $B_\phi/B_x$ expansion that controls the evolutionary spectrum properties.

Equation~\eqref{eq:exact_cov} shows that for generic noise the covariance can be dense; the band structure of Sec.~\ref{sec:offdiagonal} is a property of the noise being semi-stationary rather than of the basis.
Expressing the covariance in the frequency domain with Eq.~\eqref{eq:cov-transform},
\begin{equation}
  \Lambda_{(mn)(m'n')}=\sum_{kl}\tilde{W}_{(mn)k}\,\Sigma_{kl}\,\tilde{W}^*_{(m'n')l}\,,
  \label{eq:cov_from_Sigma}
\end{equation}
the two frequency sums are independent, and $\tilde{W}_{(mn)k}\propto\tilde{g}_{mn}(f_k)$ has support only around $\pm f_m$, Eq.~\eqref{eq:basis}.
For stationary noise, $\bm{\Sigma}$ is diagonal and the double sum collapses to a single one, so the supports of the two $\bm{\tilde{W}}$ matrices must overlap: the WDM covariance vanishes for $|m-m'|\ge2$ regardless of the spectrum shape.
For nonstationary noise $\bm{\Sigma}$ has off-diagonal terms and the two sums decouple: the WDM covariance depends on the structure of $\bm{\Sigma}$, which can be generically dense.
The tri-diagonal result relies on $B_\phi\ll B_x$, rather than being a generic property of the WDM basis.

\subsection{Discretization and data simulation}
\label{sec:cov-discrete}

To compute the covariance matrix, we discretize the frequency integral to a sum over frequency bins $f_k=k\Delta f$ with $k=0,\dots,N/2$ and $\Delta f=1/T_{\rm obs}$, weighted by
\begin{equation}
    \eta_k=
    \begin{cases}
        \tfrac{1}{2}\,, & k=0,N/2\,,\\[2pt]
        1\,, & 1\le k\le N/2-1\,,
    \end{cases}
    \label{eq:freq_edge_weight}
\end{equation}
which halves the degrees of freedom of the DC and Nyquist bins.
Equation~\eqref{eq:k_definition} then separates into the WDM transform matrix of Eq.~\eqref{eq:cov-transform} and a noise-generating matrix,
\begin{equation}
    k_{mn}(f_k)=\sum_{j=0}^{N-1}W_{(mn)j}V_{jk}\,,
    \qquad
    V_{jk}=\sqrt{\eta_k S(t_j,f_k)}\,e^{i\left[2\pi f_kt_j+\theta(t_j,f_k)\right]}\,.
    \label{eq:V_matrix}
\end{equation}
The covariance then becomes $\bm{\Lambda}=\Delta f\,\Re\left[(\mathbf{WV})(\mathbf{WV})^\dagger\right]$.
Splitting the noise generation into its real and imaginary parts, $\mathbf{V}=\mathbf{V}_c+i\mathbf{V}_s$ with
\begin{equation}
    V_{c,jk}=\sqrt{\eta_kS(t_j,f_k)}\cos\left[2\pi f_kt_j+\theta(t_j,f_k)\right]\,,
    \qquad
    V_{s,jk}=\sqrt{\eta_kS(t_j,f_k)}\sin\left[2\pi f_kt_j+\theta(t_j,f_k)\right]\,,
    \label{eq:Vcs}
\end{equation}
and using the fact that $\mathbf{W}$ is real, the covariance is
\begin{equation}
    \bm{\Lambda}=\Delta f\left[(\mathbf{W}\mathbf{V}_c)(\mathbf{W}\mathbf{V}_c)^T+(\mathbf{W}\mathbf{V}_s)(\mathbf{W}\mathbf{V}_s)^T\right]\,.
    \label{eq:wdm_generic_covariance}
\end{equation}
Data can be generated by drawing independent normal vectors $\mathbf{a}$ and $\mathbf{b}$,
\begin{equation}
    \mathbf{x}=\sqrt{\Delta f}\left(\mathbf{V}_c\mathbf{a}+\mathbf{V}_s\mathbf{b}\right)\,,
    \qquad
    x_j=\sqrt{\Delta f}\sum_{k}\sqrt{\eta_kS(t_j,f_k)}\left[a_k\cos\left(2\pi f_kt_j+\theta(t_j,f_k)\right)+b_k\sin\left(2\pi f_kt_j+\theta(t_j,f_k)\right)\right]\,.
    \label{eq:noise_realization}
\end{equation}
The WDM coefficients are $\mathbf{w}=\mathbf{W}\mathbf{x}$ and satisfy $\left\langle \mathbf{w}\mathbf{w}^T\right\rangle=\bm{\Lambda}$ by construction.
The transform matrix $\mathbf{W}$ is computed as in WDMI~\cite{Johnson:2026rrn}, including the proper normalization at the edge bands, and can be cached and reused unless the grid changes shape.

\section{Optimal choice of WDM tiling}
\label{sec:optimal}

The WDM transformation is complete, hence the likelihood is the same when calculated in either the time, frequency, or time-frequency domains as long as the covariance is estimated consistently across domains.
However, the WDM covariance is not diagonal even for stationary noise (unless the spectrum is white) and thus the likelihood can be expensive.
Approximating the WDM covariance as diagonal, $\Lambda_{(mn)(m^\prime n^\prime)} \approx 1/2 \,S_{mn} \delta_{mm^\prime}  \delta_{nn^\prime}$ where $S_{mn} = S(f=m\Delta F,t=n\Delta T)$, for computational efficiency induces some likelihood bias.

In this section, we compute the optimal WDM time-frequency tiling $(N_t, N_f=N/N_t)$ that minimizes the bias, taking into account both the time (nonstationarity) and frequency (colored noise) dependence of the power spectrum.
Intuitively, a diagonal WDM covariance matrix is achieved for stationary, white noise~\cite{Cornish:2025awt}.
Thus in order to suppress off-diagonal terms, we seek a WDM tiling with small-enough $\Delta F$ such that \emph{the spectrum is approximately flat} across the frequency support of the Meyer filter and small-enough $\Delta T$ such that \emph{the process is approximately stationary} across its temporal extent.
Both conditions are local and below we express them through the fractional change of the evolutionary spectrum $S(f,t)$ across a WDM cell.
As the WDM cell width and height are bound by $\Delta F \Delta T=1/2$, they cannot both be arbitrarily small, leading to an optimum condition that we derive here to leading order in the likelihood error.

Assuming Gaussian noise, the Kullback–Leibler (KL) divergence between the exact and diagonal likelihoods is~\cite{munoz2026kldivergencegaussiansstepbystep} 
\begin{equation}
D_\mathrm{KL}(\mathcal{N}(0,\bm{\Lambda})|| \mathcal{N}(0,\mathrm{diag} \, \bm{\Lambda})) = - \frac{1}{2} \log \det \bm{\rho}\,, 
\end{equation}
where
\begin{equation}
    \rho_{ij} \equiv \frac{\Lambda_{ij}}{\sqrt{\Lambda_{ii} \Lambda_{jj}}}\,,
    \label{eq:correlation_matrix}
\end{equation}
is the correlation matrix and $i,j$ are flattened WDM indices $(mn), (m^\prime n^\prime)$.
Separating the diagonal (which is the identity due to the normalization) and off-diagonal pieces, $\bm{\rho} \equiv \mathbf{I} + \mathbf{R}$ and expanding around the diagonal, we get
\begin{align}
\log \det \bm{\rho} &= \mathrm{Tr}\, \log \bm{\rho} \nonumber\\
&= \mathrm{Tr}\, \log (\mathbf{I} + \mathbf{R})\nonumber\\
&= \mathrm{Tr}(\mathbf R) - \frac{1}{2} \mathrm{Tr}(\mathbf{R}^2) + \mathcal{O}\left(\mathrm{Tr}(\mathbf{R}^3)\right)\,.
\end{align}
The trace of the off-diagonal piece $\mathbf R$ vanishes, so to leading order the log likelihood error is
\begin{equation}\Delta \equiv \frac{1}{4} \mathrm{Tr}\left((\bm{\rho} - \mathbf{I}) ^2\right) = \frac{1}{4}\sum_{i\neq j} \rho_{ij}^2\,.
\end{equation} 

The WDM covariance as a function of the evolutionary PSD is given in Eq.~\eqref{eq:exact_cov}.
For the purposes of the current argument, we summarize it as an operator $\mathcal{K}$ acting on the evolutionary spectrum 
\begin{align}
\Lambda_{ij} \equiv  \mathcal{K}[ S(f,t)]_{ij}\,.
\end{align}
Taylor-expanding the evolutionary PSD around $(\bar{t}, \bar{f})$ with $\bar{S}\equiv S(\bar{f},\bar{t})$, we get
\begin{align}
S(f,t) = \bar{S}\left[ 1+ \varepsilon_f\cdot(f-\bar{f})/\Delta F + \varepsilon_t \cdot(t-\bar{t})/\Delta T + \mathcal{O}(\varepsilon^2)\right]\,,
\label{eq:evspectrum-expansion}
\end{align}
where we define the dimensionless changes across the WDM cell as
\begin{align}
\varepsilon_f &= \Delta F\left(\partial_f \log S(f,t) \right)\big|_{f=\bar{f},t=\bar{t}} \,,\\
\varepsilon_t &= \Delta T\left(\partial_t \log S(f,t) \right)\big|_{f=\bar{f},t=\bar{t}} \,.
\end{align}
Then to leading order in $\varepsilon_f$ and $\varepsilon_t$,
\begin{align}
\frac{\Lambda_{ij}}{\bar{S}/2} \simeq I_{ij}+\frac{\varepsilon_f}{\Delta F} \mathcal{K}[f-\bar{f}]_{ij} + \frac{\varepsilon_t}{\Delta T} \mathcal{K}[t-\bar{t}]_{ij} + \mathcal{O}(\varepsilon^2)\,,
\end{align}
where $\mathcal{K}$ is linear to lowest order.
Defining $\mathcal{K}[f-\bar{f}]_{ij} \equiv a_{ij} \Delta F$ and $\mathcal{K}[t-\bar{t}]_{ij} \equiv b_{ij} \Delta T$, then $\rho_{ij} = a_{ij} \varepsilon_f + b_{ij} \varepsilon_t$, and the likelihood error is
\begin{align}
\Delta &= \frac{1}{4}\sum_{i\neq j} \rho_{ij}^2= \frac{1}{4} \sum_{i \neq j} \left(a_{ij}^2 \varepsilon_f^2 + 2 a_{ij} b_{ij} \varepsilon_f \varepsilon_t + b_{ij}^2 \varepsilon_t^2 \right)\Rightarrow\nonumber\\
\frac{\Delta}{N}&\equiv \frac{1}{2} \left( C_f^2 \varepsilon_f^2 + C_t^2 \varepsilon_t^2 \right)\label{eq:error-per-pixel}\,,
\end{align}
where $N$ is the number of WDM cells entering the likelihood, $N=N_tN_f$ when analyzing the full time-frequency plane or fewer when only part of it is used. 
The cross term cancels identically for the Meyer window as $a_{ij}$ and $b_{ij}$ have opposite parity under $(n-n^\prime)\to-(n-n^\prime)$, and we collapse the $a_{ij},b_{ij}$ sums into two dimensionless window-dependent constants $C_f$ and $C_t$. 
For the Meyer window used in this work $(d=4,\alpha=\Delta F / 4,\beta=\Delta F / 2)$, $C_f \approx 0.208, C_t \approx 0.639$.
Their ratio, $C_t/C_f\simeq3$, corresponds to the ratio of the extent of the Meyer window in time and frequency, expressed in units of WDM cells.
For this choice of window, a fractional change of the spectrum along the time axis induces ${\sim} 9$ times larger error than along the frequency axis.\footnote{The constants $C_t, C_f$ depend on the Meyer window parameters, with $C_f$ being a weaker function than $C_t$. For example, the Meyer window used by \citet{Pearson:2025wfd} $(d=6, \alpha=0, \beta=\Delta F)$ has $C_f=0.214, C_t = 0.498$. A 25\% lower error than our window is available at $(d=2, \alpha=0, \beta=\Delta F)$ with $C_f=0.233, C_t = 0.387$.}

Restoring the cell dependence on the derivative terms, the log likelihood error in terms of $N_t$ (as $N_f=N/N_t$) is
\begin{align}
\Delta(N_t) = \frac{N}{2} \left[ \frac{C_f^2 \left( \partial_f \log S \right)^2}{4 T_\mathrm{obs}^2} N_t^2 + C_t^2 \left(\partial_t \log S\right)^2 T_\mathrm{obs}^2 \frac{1}{N_t^2} \right]\,,
\label{eq:error-Nt}
\end{align}
which has a unique minimum,
\begin{align}
N_t^\star &= T_\mathrm{obs}\sqrt{\frac{2C_t |\partial_t \log S|}{C_f |\partial_f \log S|}}\,,\label{eq:optimal_tiling}\\
\frac{\Delta^\star}{N} &= \frac{C_f C_t}{2} \left|\partial_f \log S \cdot \partial_t \log S \right|\,.
\label{eq:optimal_loss}
\end{align}

In practice, PSDs rarely have constant derivatives, so Eq.~\eqref{eq:optimal_tiling} will give a range of values of $N_t^\star$. 
In that case, we use the RMS derivatives averaged over the time-frequency region entering the likelihood:
\begin{align}
    N_t^\star &= T_\mathrm{obs}\sqrt{\frac{2C_t}{C_f}\frac{\langle (\partial_t \log S)^2 \rangle^{1/2}}{\langle (\partial_f \log S)^2 \rangle^{1/2}}}\,,\label{eq:rms_optimal_tiling}\\
    \frac{\Delta^\star}{N} &= \frac{C_f C_t}{2} \langle (\partial_f \log S)^2 \rangle^{1/2} \cdot \langle (\partial_t \log S)^2 \rangle^{1/2}\,.
\label{eq:rms_optimal_loss}
\end{align}
For a given PSD (and data duration), these equations provide the optimal tiling and the log likelihood error to leading order in the PSD dependence, $\varepsilon_f$ and $\varepsilon_t$.
In Eq.~\eqref{eq:rms_optimal_loss}, the product of the time and frequency derivatives provides the smallest possible error, while in Eq.~\eqref{eq:rms_optimal_tiling} their ratio determines the optimal tiling.
At the optimum, the two terms of Eq.~\eqref{eq:error-Nt}, corresponding to deviation from whiteness and nonstationarity respectively, contribute equally to the error.

The log likelihood error in Eq.~\eqref{eq:rms_optimal_loss} can determine whether the diagonal covariance is an admissible approximation.
The KL divergence $\Delta$ corresponds to the expected difference in the log likelihood when analyzing the data with the exact vs the diagonal covariance. 
A threshold of $\Delta^\star \lesssim 1$ corresponds to a log Bayes factor between the two noise models of $1$, i.e., roughly a distinguishability threshold.
The criterion is based on differences in noise modeling and thus is more directly applicable to noise inference; it is conservative when applied to signal inference. 
This is because $\Delta$ corresponds to the error in the value of the log likelihood over the entire time-frequency space of $N$ cells.
Signal inference, on the other hand, depends on differences in log likelihood over the time-frequency cells occupied by the signal (and its prior).
Moreover, $\Delta$ scales with $N$, the number of WDM cells entering the likelihood. 
In the context of LISA, for example, off-diagonal terms are more costly over the entire global fit than a narrow band around a single resolved Galactic binary, see Sec.~\ref{sec:result-tiling}.

\section{Welch-like coarse-graining}
\label{sec:coarse-graining}

While LISA inference will likely model the evolutionary PSD and infer it using a likelihood, the PSD evaluated directly from the data might have applications in stochastic background inference, for example to accelerate likelihood evaluations~\cite{Caprini:2019pxz,Flauger:2020qyi,Baghi:2026dfk}.
In the frequency domain, PSD estimation based on a realization of the data relies on Welch's~\cite{Welch:1967} and Bartlett's~\cite{Bartlett:1948} methods.
The data are divided into several overlapping (Welch) or nonoverlapping (Bartlett) windowed time segments.
The PSD of each segment is estimated independently, via the amplitude of the Fourier components, and then averaged across segments.
Compared to the PSD of the full dataset with resolution $1/T_{\rm obs}$, a Welch PSD from $N_\mathrm{seg}$ segments has a coarser resolution at $N_\mathrm{seg}/T_{\rm obs}$, and the likelihood is thus faster to evaluate.
It also has different noise statistics: for stationary Gaussian noise, each segment's PSD is $\chi^2$-distributed with an effective number of degrees of freedom~\cite{Baghi:2026dfk} and the average decreases the variance of the PSD estimate proportionally to $N_\mathrm{seg}$.

In this section, we consider such PSD coarse-graining in the WDM domain across the time-frequency grid, without the assumption of stationarity.
The need for PSD averaging, at least along the time axis, was already discussed in Sec.~\ref{sec:evolutionary-estimation}.
\citet{Priestley1966} showed that the nonaveraged PSD has 100\% variance independently of $T_{\rm obs}$, while time-averaging reduces the variance by a factor proportional to the number of filters in the data averaged over, $Q=T_q/B_\phi$ with $B_\phi \simeq \Delta T$ for the WDM Meyer filter.
We extend to frequency averaging, relax the diagonality assumption, and compute the nonstationarity correction for time averaging.

We consider nonstationary, colored Gaussian noise in the WDM domain, where the WDM coefficients are distributed as $\mathbf{w} \sim \mathcal{N}(0,\bm{\Lambda})$, where $\bm{\Lambda}$ is the WDM covariance given in Eq.~\eqref{eq:offdiag_covariance} to leading order in $B_\phi/B_x$ or in Eq.~\eqref{eq:wdm_generic_covariance} exactly.
We divide the WDM plane into non-overlapping time-frequency regions $R_\mu$, each containing $|R_\mu|$ cells. 
The regions need not be equal-area nor square.
We then define the coarse-grained WDM spectrum of region $\mu$ by averaging the WDM power across the region,
\begin{align}
    P_{\mu} \equiv \frac{1}{|R_\mu|} \sum_{(mn)\in R_\mu} w_{mn}^2\,.
\end{align}
Since the WDM coefficients are normally distributed with covariance $\Lambda_{(mn)(m'n')}$, $P_{\mu}$ is a sum of squares of correlated Gaussian random variables.
Quadratic Gaussian forms like $P_{\mu}$  have known cumulant (mean, variance, etc.) expansions \cite{lancasterQuadraticForms}.
Denoting the block of $\bm{\Lambda}$ corresponding to region $R_\mu$ by $\mathbf{B}_\mu$, that is the $|R_\mu|\times|R_\mu|$ matrix with entries $(\mathbf{B}_\mu)_{(mn)(m'n')}=\Lambda_{(mn)(m'n')}$ for $(mn),(m'n')\in R_\mu$, then the $s$-th cumulant of $P_\mu$ is
\begin{align}
    \kappa_{s,\mu} = \frac{2^{s-1}(s-1)!}{|R_\mu|^s} \,\mathrm{Tr}(\mathbf{B}_\mu^s)\,.
    \label{eq:coarse-grained-cumulants}
\end{align}
While the resulting distribution does not generally have a closed form, we approximate it with a Gamma distribution by matching the first two cumulants with the Welch-Satterthwaite approximation~\cite{Satterthwaite:1946,WelchBL:1947,Box:1954}:
\begin{align}
    \label{eq:coarse-grained-WS}
    P_\mu &\sim \Gamma(\alpha_\mu, \theta_\mu)\,, \qquad \text{with}\quad\alpha_\mu \equiv \kappa_{1,\mu}^2 / \kappa_{2,\mu}\,,\,\, 
    \theta_\mu \equiv \kappa_{2,\mu} / \kappa_{1,\mu}\,.
\end{align}
The approximation is exact when all eigenvalues of $\mathbf{B}_\mu$ are equal. 
Its error can be estimated comparing the skewness of the Gamma distribution to the true skewness from Eq.~\eqref{eq:coarse-grained-cumulants}. 
An infinite series description of the exact distribution also exists, although in our tests it converged too slowly to be practically useful~\cite{MoschopoulosP.G.1985Tdot}.

Besides correlations between the WDM cells within a region, that Eq.~\eqref{eq:coarse-grained-WS} takes into account, correlations exterior to the region also persist after coarse-graining, $\mathrm{Cov}(P_\mu,P_\nu) \neq 0$. 
To estimate them, we start with the covariance between two squared WDM cells
\begin{align}
    \mathrm{Cov}(w_{mn}^2,w_{m'n'}^2) &\equiv \langle w_{mn}^2w_{m'n'}^2\rangle - \langle w_{mn}^2\rangle \langle w_{m'n'}^2\rangle\nonumber\\
    &= \langle w_{mn}^2\rangle\langle w_{m'n'}^2\rangle + 2 \langle w_{mn} w_{m'n'}\rangle^2 - \langle w_{mn}^2\rangle \langle w_{m'n'}^2\rangle\nonumber \\
    &= 2 \Lambda_{(mn)(m'n')}^2\,,
\end{align}
where in the middle line we have expanded the fourth moment using Isserlis's Theorem \cite{isserlis1918}.
The covariance between coarse-grained regions is then
\begin{align}
    \mathrm{Cov}(P_\mu,P_\nu) &= \frac{1}{|R_\mu||R_\nu|} \sum_{(mn)\in R_\mu} \sum_{(m'n')\in R_\nu} \mathrm{Cov}(w_{mn}^2, w_{m'n'}^2) \nonumber \\
    &=\frac{2}{|R_\mu||R_\nu|} \sum_{(mn)\in R_\mu} \sum_{(m'n')\in R_\nu} \Lambda_{(mn)(m'n')}^2\,.
    \label{eq:coarse-grained-correlations}
\end{align}

The utility of coarse-graining lies in Eq.~\eqref{eq:coarse-grained-correlations}.
Averaging the squared coefficient of some region into $P_\mu$ reduces the numbers entering the likelihood by a factor of $|R_\mu|$ and replaces the distribution from Gaussian in $w_{mn}$ to a Gamma function in $P_\mu$.
While the intra-region correlations are incorporated in Eq~\eqref{eq:coarse-grained-cumulants}, Eq.~\eqref{eq:coarse-grained-correlations} shows that the cross-region correlations are \textit{suppressed quadratically by coarse-graining.}
Concretely, if the WDM correlation coefficient is $\rho$, Eq.~\eqref{eq:correlation_matrix}, the correlation coefficient of their squares is $\rho^2$.  
In principle, if one were to condense the entire time-frequency plane into a single coarse-grained region, all covariances would be absorbed. 
But this would come at the cost of time and frequency resolution.
A more optimal strategy might coarse-grain high-correlation regions to absorb large internal correlations while suppressing cross-region correlations with Eq.~\eqref{eq:coarse-grained-correlations}.  
We leave a full exploration of optimizing coarse-graining for noise and stochastic background inference to future work.

\subsection{Time coarse-graining}

As a special case, similar to Welch's method and Priestley's averaging, we consider coarse-graining purely in the time direction. 
We time-average over $Q$ WDM cells, that evenly divide $N_t$, for a coarse-graining window $T_q=Q\Delta T$. 
The coarse-grained power is then obtained averaging the coefficients squared of $Q$ cells,
\begin{align}
    P_{mq} \equiv \frac{1}{Q} \sum_{n=qQ}^{(q+1)Q-1} w_{mn}^2\,,
\end{align}
which reduces the time resolution by a factor of $Q$.
The corresponding covariance block is $\mathbf{B}_{mq}$, of size $Q\times Q$, with entries $(\mathbf{B}_{mq})_{nn'}=\Lambda_{(mn)(mn')}$ at fixed band $m$ and $n,n'\in[qQ,(q+1)Q)$.
This equation is the discretized version of Priestley's time-averaging in Eq.~\eqref{eq:time-smoothing} evaluated on the WDM grid and for a boxcar window of length $T_q$.

If the noise is approximately stationary over the $Q$ cells, $T_q\ll B_x$, and the cells are uncorrelated, then $\mathbf{B}_{mq}$ is proportional to the identity and the Welch-Satterthwaite matching is exact,
\begin{equation}
P_{mq}\sim \Gamma\left(\frac{Q}{2}, \frac{2\bar{S}_{2,mq}}{Q}\right)\,,
\end{equation}
where $\bar{S}_{2,mq} \equiv \mathrm{Tr}(\mathbf{B}_{mq})/Q$ is the time-averaged PSD: the region-averaged diagonal of the covariance, that is $\varepsilon_{mn}S_2(t_n,f_m)$ of Eq.~\eqref{eq:wdm_variance_cell_center} averaged over the region.
The first Gamma argument follows from Eq.~\eqref{eq:coarse-grained-WS} and counts the $Q$ real degrees of freedom of the region (rather than the two per bin in the Fourier domain).
The mean and variance of the above distribution are 
\begin{equation}
    \langle P_{mq}\rangle = \bar S_{2,mq}\,,
    \qquad
    \mathrm{Var}[P_{mq}] = \frac{Q}{2}\left(\frac{2\bar S_{2,mq}}{Q}\right)^2
    = \frac{2\,\bar S_{2,mq}^2}{Q}=\frac{2}{Q}\left(\langle P_{mq}\rangle\right)^2\,.
\end{equation}
The latter is the discrete version of Priestley's smoothed variance in Eq.~\eqref{eq:smoothed-variance}, since the integral term of Eq.~\eqref{eq:smoothed-variance} for the boxcar filter is $1/\Delta F$ and $T_q \Delta F= Q/2$.

Beyond stationary noise, for a within-region varying PSD the Welch-Satterthwaite approximation implies an effective shift \begin{equation}
    Q\rightarrow Q_{\mathrm{eff},mq} = \frac{\mathrm{Tr}(\mathbf{B}_{mq})^2}{\mathrm{Tr}(\mathbf{B}_{mq}^2)}\simeq\frac{\left(\sum_{n=qQ}^{(q+1)Q-1}\varepsilon_{mn}S_{mn}\right)^2 }{ \sum_{n=qQ}^{(q+1)Q-1}\varepsilon_{mn}S_{mn}^2}\,,
    \label{eq:Qeff-def}
\end{equation} 
where the second equation follows for independent WDM cells.
The effective $Q$ shift quantifies the impact of the assumption that the spectrum is constant across the region.
Decomposing it into a region-averaged term and a time variation, $S_{mn}=\bar{S}_m(1+\delta_n)$, where the variation has zero mean $\langle\delta_n\rangle=0$ and fractional variance $\sigma^2_\delta$, Eq.~\eqref{eq:Qeff-def} becomes
\begin{equation}
Q_{\mathrm{eff},mq}=\frac{Q}{1+\sigma^2_\delta}\simeq Q\left[1-\sigma^2_\delta\right]\,.
    \label{eq:Qeff_drift}
\end{equation}
For a linear drift of fractional amplitude $T_q/B_x$ across the region, $\sigma^2_\delta=\left(T_q/B_x\right)^2/12$.
This equation provides the leading-order correction to the $T_{q}\ll B_x$ assumption of Sec.~\ref{sec:evolutionary-estimation} and shows that it is a second-order effect.

For the exact joint likelihood of coarse-grained data we would need to take into account all between-region covariances and higher order moments beyond the covariance.
However, since coarse-graining reduces covariances, the joint Welch-Satterthwaite likelihood can approximately neglect all between-region correlations
\begin{align}
    \mathcal{L} \simeq \prod_{m=0}^{N_f} \prod_{q=0}^{N_t/Q-1} \Gamma\left(P_{mq}\,;\frac{Q_{\mathrm{eff},mq}}{2}, \frac{2 \bar{S}_{2,mq}}{Q_{\mathrm{eff},mq}} \right)\,.
\end{align}
The accuracy of this approximation can be checked via Eq.~\eqref{eq:coarse-grained-correlations} or by comparison to more computationally expensive, non-coarse-grained inference on the same data.
Assuming the evolutionary PSD is roughly known and $Q_\mathrm{eff}$ is precomputed, coarse-graining in time accelerates the likelihood evaluation by a factor of $Q$ relative to a diagonal Gaussian likelihood, while reducing covariances.

\begin{figure}[t]
     \centering
     \includegraphics[width=\columnwidth]{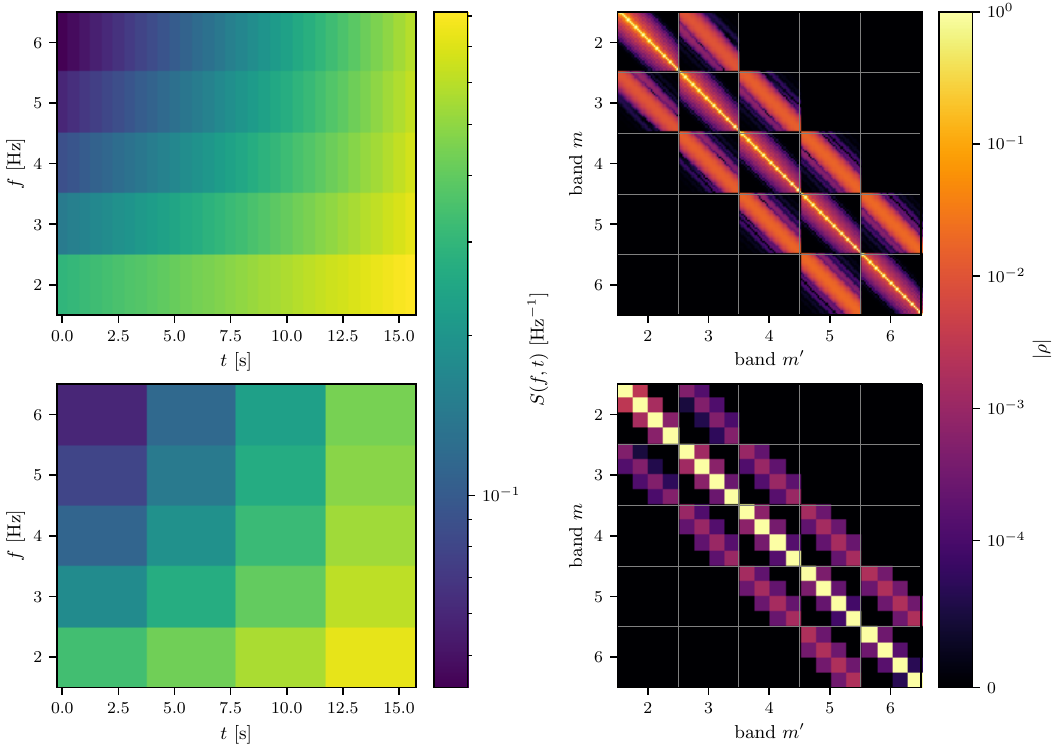}
     \caption{Comparison between the evolutionary PSD of Sec.~\ref{sec:monochoromatic-methods} (index-drifting powerlaw) (left column) and its correlation matrix $\bm{\rho}$ from Eq.~\eqref{eq:correlation_matrix} (right column) at full sampling with $N_f=8, N_t=32, \Delta t = 1/16 \,\mathrm{s}$ (top row) and their time-coarse-grained counterparts with $Q=8$ (bottom row). We use an observation band restricted to $f\in[2,6]\,
     \mathrm{Hz}$. The maximum off-diagonal correlation element with full sampling is $\approx 0.18$, while coarse-graining reduces it to $\approx 0.0029$.}
     \label{fig:coarse-graining}
\end{figure}

 In Figure~\ref{fig:coarse-graining}, we apply coarse-graining to the evolutionary PSD for the index-drifting powerlaw noise described in Sec.~\ref{sec:monochoromatic-methods}. 
 By coarse-graining in the time direction, correlations are significantly reduced while also reducing the amount of data entering the inner products and likelihood.

\section{Inference Methods}
\label{sec:methods}

To illustrate the role of the covariance and its approximations, we consider simulated data with Gaussian noise. 
In this section, we present the relevant methods.
We explore two simulated cases: a monochromatic signal in powerlaw noise whose spectral index varies over time, Sec.~\ref{sec:monochoromatic-methods} and LISA data with a single resolvable Galactic binary with subthreshold binaries creating cyclostationary noise, Sec.~\ref{sec:UCB-methods}.
We analyze each example with different covariances and likelihoods, detailed in Sec.~\ref{sec:likelihoodmethod}, with implementation details given in Sec.~\ref{sec:implementation}.

\subsection{Likelihood and SNR definitions}
\label{sec:likelihoodmethod}

We generate data $d$ with signal $h$ and noise $n$ such that $d = n + h$.
The noise is Gaussian but not stationary, instead generated using the full covariance matrix of Sec.~\ref{sec:cov-generation} to all orders in the ratio of the filter to nonstationarity timescales.
For inference, we use a Gaussian likelihood in which the domain and covariance treatment varies.
Specifically, we consider:
\begin{enumerate}
\item A frequency-domain Whittle (stationary) likelihood with diagonal covariance matrix, i.e., Eq.~(116) in WDMI. 
This corresponds to the stationary noise assumption, where the PSD $S(f)$ is modeled and inferred as described below.
This likelihood effectively ``averages" over the time dependence of the noise and provides a benchmark for the bias induced by ignoring nonstationarity.
\item A WDM likelihood with the exact covariance matrix, i.e., Eq.~\eqref{eq:wdm-gaussian} with Eq.~\eqref{eq:wdm_generic_covariance} for the covariance. The evolutionary PSD $S(t,f)$ is modeled and inferred.
This covariance is not formally accessible for inference on real data as it depends on the time dependence of the oscillatory envelope phase, $\theta(t,f)$. 
Here we evaluate it on the simulated value of $\theta(t,f)$ only for comparison as it provides the exact posterior.
This is the most computationally expensive likelihood as the covariance matrix is generically dense.
\item A WDM likelihood with the tri-diagonal covariance matrix arising from the timescale assumption that the noise varies slowly compared to the filter extent, i.e., Eq.~\eqref{eq:wdm-gaussian} with Eq.~\eqref{eq:offdiag_covariance} for the covariance. The evolutionary PSD $S(t,f)$ is again modeled and inferred.
This likelihood models the time-variation of the noise to leading order and it is the most complete covariance that can be computed for real data without additional assumptions about the noise properties.
\item A band-averaged diagonal WDM likelihood. As the full and tri-diagonal WDM likelihoods are expensive to evaluate, we consider the diagonal approximation corresponding to Eq.~(121) in WDMI but where the diagonal elements of the covariance are given by Eq.~\eqref{eq:wdm_variance_band_average}.
The evolutionary PSD $S(t,f)$ is again modeled and inferred.
The error induced by this approximation is calculated to leading order in Sec.~\ref{sec:optimal}.
\item A midpoint diagonal WDM likelihood. This is again diagonal per Eq.~(121) in WDMI but now the evolutionary PSD is evaluated at the WDM cell's midpoint rather than averaged by the square of the Meyer filter, Eq.~\eqref{eq:wdm_variance_cell_center} and also Eq.~(118) in WDMI. 
\end{enumerate}
We do not present results with a WDM stationary likelihood as it is equivalent to the frequency Whittle one since the transform is complete.
We have confirmed this numerically using simulated stationary noise and evaluating the WDM likelihood with the full stationary tri-diagonal covariance.

For the exact WDM likelihood, the choice of WDM tiling does not matter: the transformation is complete so any choice of $\Delta T$ and $\Delta F=1/(2\Delta T)$ will yield the same likelihood.
This is not the case for the tri-diagonal and diagonal versions, which are approximate and therefore tiling-dependent, as quantified in Sec.~\ref{sec:optimal}.
The band-averaged likelihood is optimal for the tiling that minimizes off-diagonal terms, while the midpoint likelihood further favors a tiling for which $S(t, f)$ is nearly constant across the support of the Meyer filter in both frequency and time.
The formalism of Sec.~\ref{sec:optimal} for the optimal tiling and the expected likelihood error from the diagonal approximation is applied to the LISA noise in Sec.~\ref{sec:result-tiling}.

\begin{table}[!tbp]
    \centering
    \begin{tabular}{l|cccc}
        {Covariance} & {raw} & {cut} & {exact edges} & {padded} \\\hline
        {Whittle (stationary)}                  & {$\bullet$} & {$\bullet$} & {--} & {--} \\
        {Exact}                                 & {$\bullet$} & {--} & {--} & {--} \\
        {Tri-diagonal}                          & {$\bullet$} & {$\bullet$} & {$\bullet$} & {$\bullet$}\\
        {Band-averaged diagonal}                & {$\bullet$} & {$\bullet$} & {$\bullet$} & {$\bullet$} \\
        {Midpoint diagonal}                     & {$\bullet$} & {$\bullet$} & {$\bullet$} & {$\bullet$} 
    \end{tabular}
    \caption{The likelihoods and their corresponding covariances we explore for each simulation. Every approximate covariance is combined with a time-edge treatment to mitigate the larger correlations due to data wraparound.}
    \label{tab:likelihood-matrix}
\end{table}

Like the discrete Fourier transform, the WDM transform treats any finite-length data as periodic at the time boundaries, $w_{mn} = w_{m(n+N_t)}$.
This means that the Meyer filter of the time-edge bins ``sees" the data on the other end, adding a correlation between the time-frequency edge cells.
Practically speaking, if the noise level is different between the start and end of the data, there is a considerable discrepancy between the start and end values and the corresponding correlation can be substantial.
For example, in the case of the drifting powerlaw noise of Sec.~\ref{sec:monochoromatic-methods}, the spectrum ratio between the start and end of the observed data for the 137~Hz signal is 366.
The exact WDM covariance of Eq.~\eqref{eq:wdm_generic_covariance} already captures this periodicity and corresponding correlations through the periodicity of the transform matrix $\mathbf{W}$.
The large correlations induced by wraparound are, however, ignored in the tri-diagonal and diagonal covariances, leading to a large error.
We consider the following strategies, summarized in Table~\ref{tab:likelihood-matrix}:
\begin{enumerate}
    \item No mitigation.
    \item Excising a few time-edge columns of the time-frequency grid, with $n$ values close to $0$ (left edge) and $N_t-1$ (right edge).
    Specifically, we cut every column for which the time-domain Meyer filter squared falls to 0.1\% of its maximum value after it has wrapped around to the other side, computed numerically.
    For the Whittle likelihood, we crop the data to the time span of the columns kept.
    \item 
    Replacing a set of time-frequency columns with the exact covariance.
    This is again for comparison only as we realistically do not have access to the exact covariance.
    Where the resulting covariance is not positive definite, the likelihood returns $-\infty$.
    \item Zero-padding the data and the template.   
    We transform on the extended grid keeping $N_t$ columns, and replace the covariance by $\mathbf{M}\boldsymbol{\Lambda}\mathbf{M}$, where $\mathbf{M}$ is the data mask, see Sec.~\ref{sec:implementation}.
\end{enumerate}

Finally, the different likelihoods and covariance approximations have their own corresponding signal-to-noise ratios (SNRs).
This situation is similar to the tiling discussed above: while the estimates should agree between the Fourier and WDM domains for the exact covariance matrix, approximating either domain's covariance changes the SNR.
We denote each likelihood's SNR respectively as (1) $\rho_\mathrm{whittle}$, (2) $\rho_\mathrm{exact}$, (3) $\rho_\mathrm{tri}$, (4) $\rho_\mathrm{band}$, (5) $\rho_\mathrm{mid}$.
In the frequency-domain the SNR is
\begin{equation}
    \rho_\mathrm{whittle}^2 = \langle h \mid h\rangle = 4\sum_{k=0}^{N/2} \eta_k\frac{|h(f_k)|^2}{S(f_k)}\Delta f\,,
    \label{eq:whittle_snr}
\end{equation}
where $\eta_k$ is the edge weight of Eq.~\eqref{eq:freq_edge_weight}, and $f_k = k \Delta f$.
In the WDM domain, the SNR is
\begin{equation}
    \rho_\mathrm{WDM}^2 = \mathbf{h}^T\bm{\Lambda}^{-1}\mathbf{h}\,,
\end{equation}
but using different data (when excised) or covariance matrix from the choices above: exact, tri-diagonal, band-averaged, or midpoint.

\subsection{Monochromatic signal in drifting noise}
\label{sec:monochoromatic-methods}

\begin{table}[!tbp]
    \centering
    \begin{tabular}{l|c|c}
        Parameter & Simulation & Prior \\
        \hline
        \multicolumn{3}{l}{\textit{Signal}}\\
        Amplitude $A_h$                                   & \{$1.632, 0.3042, 0.08730$\}    & U[$0$, $3$]\\
        Frequency $f_0$ [Hz]                              & $\{33, 137, 385\}$    & U[16, 496] \\
        Phase $\phi_0$ [rad]                              & $0.6$    & U[$-\pi$, $\pi$] \\\hline
        \multicolumn{3}{l}{\textit{Noise}}\\
        Amplitude $\log_{10}(A_{n}/\mathrm{Hz^{-1}})$     & $2.602$  & U[$1$, $4$] \\
        Spectral index intercept $\alpha_c$               & $-2.0$   & U[$-4$, $0$] \\
        Spectral index slope $\alpha_d$                   & $1.2$    & U[$-2$, $2$] \\
    \end{tabular}
    \caption{Parameters for the simulated monochromatic signal in drifting powerlaw Gaussian noise of Sec.~\ref{sec:monochoromatic-methods}. We list the true simulated values for the signal and noise parameters as well as the uniform sampling prior, with lower and upper bounds in square brackets. The powerlaw drift parameter, $\alpha_d$, only enters the WDM likelihood; the frequency-domain likelihood instead assumes a constant spectral index, parametrized by $\alpha_c$. }
    \label{tab:sine_params}
\end{table}

We first consider a simple toy model: a pure sinusoid signal in Gaussian noise with a drifting powerlaw PSD.
We simulate $T_{\mathrm{obs}}=16\,$s of data sampled at $\Delta t=1/1024\,$s for a total of $N=T_{\mathrm{obs}}/\Delta t = 16384$ data points.
The sampling frequency is $f_s = 1024\,$Hz and Nyquist is $f_s/2 = 512\,$Hz.
These parameters determine the frequency-domain likelihood settings.
In the WDM domain, we select cells with width $\Delta T = T_{\mathrm{obs}}/N_t = 1/16\,$s and height $\Delta F = 1/(2\Delta T) = 8\,$Hz.
The frequency axis is tiled with $N_f = 64$ and the time axis with $N_t = 256$ cells, confirming $N_f N_t = N = 16384$.
In all domains, the minimum analysis frequency is $f_{\rm min}=4\,$Hz.

In the time domain, the signal is
\begin{equation}
h(t)= A_{h} \cos (2\pi f_0 t+\phi_0)\,,
\label{eq:sine_example}
\end{equation}
with parameters the amplitude $A_h$, frequency $f_0$, and phase $\phi_0$. 
The noise PSD is a powerlaw spectrum with a spectral index that drifts linearly in time
\begin{equation}
S(t, f) = A_{n} \left(\frac{\max(f,f_\mathrm{b})}{f_{\mathrm{ref}}}\right)^{\alpha(t)}\,, \qquad \alpha(t) = \alpha_c + \alpha_d \left(\frac{t}{T_{\mathrm{obs}}}-\frac{1}{2}\right)\,,
\label{eq:sine_example_psd}
\end{equation}
with parameters including the noise amplitude $A_n$, spectral index at the midpoint of the data $\alpha_c$, and slope $\alpha_d$.
Setting $\alpha_d=0$ returns a stationary powerlaw with spectral index $\alpha_c$.
We impose a minimum frequency of $f_\mathrm{b}=4\,$Hz to keep the spectrum finite and the covariance matrix well-conditioned and set $f_{\mathrm{ref}}=1\,$Hz.
The simulated values for the signal and noise parameters are given in Table~\ref{tab:sine_params}.
Across the observation, the noise spectrum drifts from $\alpha(t=0)=-2.6$ to $\alpha(t=T_{\mathrm{obs}})=-1.4$.
Thus the PSD shape changes over time, with some parts of the spectrum shifting more than others.
The error of approximating the covariance depends on the logarithmic derivatives of $S(t,f)$ with time and frequency, Eq.~\eqref{eq:error-per-pixel}.
The two terms have opposite effects across the spectrum: the time drift grows with frequency, $\varepsilon_t\propto\log(f/f_\mathrm{ref})$, while the spectral term, $\varepsilon_f\propto1/f$, is higher at low frequencies.
To probe the error budget, we consider signals with $f_0=\{33, 137, 385\}\,$Hz for which $(C_f \varepsilon_f)^2/(C_t \varepsilon_t)^2 = 93, 2.7$, and $0.23$ respectively.
The diagonal approximation is limited by the spectral slope at $33\,$Hz, by both at $137\,$Hz, and by the time drift at $385\,$Hz.
Each signal is simulated with SNR $\rho_{\mathrm{exact}} = 15$ (minimally reduced by the 4~Hz cut). 

All inference described in Sec.~\ref{sec:likelihoodmethod} models the signal through Eq.~\eqref{eq:sine_example}, but makes different assumptions about the noise model.
The frequency-domain Whittle likelihood includes only the noise spectrum amplitude, and (fixed in time) spectral index.
In contrast, every WDM likelihood uses the full Eq.~\eqref{eq:sine_example_psd} relaxing the assumption of stationary noise and modeling the noise spectrum as a powerlaw whose spectral index has a slope and an intercept.
The wraparound mitigation involves excising the first 6 and last 5 time-edge bins, replacing the first 8 and the last 7 bins with the exact covariance, or adding 8 zero-padded bins on each side.
Priors for all parameters are given in Table~\ref{tab:sine_params}.

\subsection{LISA Galactic binary in cyclostationary noise}
\label{sec:UCB-methods}

\begin{table}[!tbp]
    \centering
    \begin{tabular}{l|c|c|c|c|c|c}
        Parameter & \multicolumn{2}{c|}{$3$ months} & \multicolumn{2}{c|}{$6$ months} & \multicolumn{2}{c}{$1$ year} \\
                                   & $A$ & $E$ & $A$ & $E$ & $A$ & $E$ \\\hline
        \multicolumn{7}{l}{\textit{Instrument, held fixed}}\\
        $S_{\rm oms,d}$ [m$^2$\,Hz$^{-1}$] & \multicolumn{6}{c}{$6.241\times10^{-23}$} \\
        $S_{a}$ [m$^2$\,s$^{-4}$\,Hz$^{-1}$] & \multicolumn{6}{c}{$5.76\times10^{-30}$}  \\
        $f_*$ [mHz]      & \multicolumn{6}{c}{$19.09$}               \\\hline
        \multicolumn{7}{l}{\textit{Foreground spectrum $S_{\rm fg}(f)$, shared across channels}}\\
        $\tilde{A}_{\rm fg}$ [Hz$^{-1}$]     & \multicolumn{2}{c|}{$1.158\times10^{-40}$} & \multicolumn{2}{c|}{$1.302\times10^{-40}$} & \multicolumn{2}{c}{$1.728\times10^{-40}$} \\
        $\alpha$                  & \multicolumn{2}{c|}{$1.716$}  & \multicolumn{2}{c|}{$1.713$}  & \multicolumn{2}{c}{$0.923$}  \\
        $\tilde{s}_1$       & \multicolumn{2}{c|}{$24.13$}  & \multicolumn{2}{c|}{$35.18$}  & \multicolumn{2}{c}{$8.972$}  \\
        $\tilde{s}_2$       & \multicolumn{2}{c|}{$1909$}  & \multicolumn{2}{c|}{$43.77$}  & \multicolumn{2}{c}{$39.92$}  \\
        $x_k$               & \multicolumn{2}{c|}{$0.2093$}  & \multicolumn{2}{c|}{$0.2027$}  & \multicolumn{2}{c}{$0.1486$}  \\\hline
        \multicolumn{7}{l}{\textit{Power modulation $\log g_c(t)$, per channel}}\\
        {$a_{c,0}$} & {$-7.380$} & {$-2.641$} & {$-0.948$} & {$0.636$} & {$-0.039$} & {$0.072$} \\
        {$a_{c,1}$} & {$5.177$} & {$2.656$} & {$-0.480$} & {$-0.397$} & {$-0.293$} & {$-0.366$} \\
        {$b_{c,1}$} & {$8.298$} & {$2.363$} & {$1.232$} & {$-0.951$} & {$-0.136$} & {$-0.024$} \\
        {$a_{c,2}$} & {$0.552$} & {$-0.691$} & {$0.047$} & {$-0.886$} & {$-0.550$} & {$-0.485$} \\
        {$b_{c,2}$} & {$-2.282$} & {$-1.116$} & {$0.146$} & {$-0.047$} & {$-0.031$} & {$-0.055$} \\
    \end{tabular}
    \caption{Noise parameters for the LISA simulations of Sec.~\ref{sec:UCB-methods}.
    The instrument parameters are held fixed throughout. 
    The foreground spectrum is shared across TDI channels, while the power modulation is distinct per channel $c\in\{A,E\}$. 
    The foreground spectrum and its modulation are fitted to the signal-free data for each of the three observation durations.}
    \label{tab:foreground-coeffs}
\end{table}

The second simulation is of LISA data containing a single resolvable Galactic binary, a set of unresolvable binaries from the LDC2a dataset~\cite{baghi2022lisadatachallenges}, and Gaussian instrument noise.
The unresolvable signals sum to an approximately Gaussian foreground with an annual power modulation as the LISA antenna pattern points toward and away from the center of the Galaxy.
We consider different observation durations $T_{\rm obs}$ to probe a range of noise variations, from drifting (over a few months) to cyclostationary (over a year). 
We use first-generation Michelson combinations to form the $X$, $Y$, and $Z$ time-delay interferometry (TDI) channels, which we then combine into $A$ and $E$.
We assume that $A$ and $E$ are orthogonal and avoid the $T$ channel.

As we do not have an exact model for the foreground, we instead model the noise in each LISA channel $c\in\{A,E\}$ as Gaussian with an evolutionary spectrum 
\begin{equation}
    S_c(t, f) = S_\mathrm{inst}(f) + g_c(t)S_\mathrm{fg}(f)\,,
    \label{eq:LISA-noise-model}
\end{equation}
where $g_c(t) > 0$ is the power modulation, $S_\mathrm{fg}(f)$ is the foreground spectrum, and $S_\mathrm{inst}(f)$ is the instrument PSD; the latter two are shared across channels.
The instrument PSD is fixed to its LDC2a form throughout~\cite{Robson:2018ifk,baghi2022lisadatachallenges}
\begin{align}
    S_{\rm inst}(f)&=8\sin^2\!x\,\Big[S_{\rm oms}(f)\,(2+\cos x)+2S_{\rm acc}(f)\,(3+2\cos x+\cos2x)\Big]\,,\label{eq:LISA-instrument}\\
  S_{\rm oms}(f)&=S_{\rm oms,d}\Big[1+\big(\tfrac{2\,\mathrm{mHz}}{f}\big)^4\Big]\Big(\tfrac{2\pi f}{c_0}\Big)^2\,,\\
  S_{\rm acc}(f)&=S_a\Big[1+\big(\tfrac{0.4\,\mathrm{mHz}}{f}\big)^2\Big]\Big[1+\big(\tfrac{f}{8\,\mathrm{mHz}}\big)^4\Big]\frac{1}{(2\pi f)^2c_0^2}\,,
\end{align}
with $x\equiv f/f_*$ the dimensionless transfer frequency, and $c_0$ the speed of light.
Parameter values are given in Table~\ref{tab:foreground-coeffs}.
The foreground and its modulation are modeled as~\cite{Katz:2024oqg}
 \begin{align}
  S_{\rm fg}(f)&=3 \tilde{A}_{\rm fg}\,x^{-1/3}\sin^2\!x\;e^{-\tilde{s}_1 x^{\alpha}}\,\big[1+\tanh\left(-\tilde{s}_2(x-x_k)\right)\big]\,,\label{eq:foreground_fit}\\
  \log g_{c}(t) &= a_{c,0} + \sum_{r=1}^{2}\left[a_{c,r} \cos(r\Omega t)+b_{c,r}\sin(r\Omega t)\right]\,,\label{eq:foreground_modulation}
\end{align}
with $\Omega = 2\pi/T_{\mathrm{yr}}$ and $T_\mathrm{yr} = 365\,\mathrm{d}$.
The remaining parameters are fitted from the data as described below.
In Priestley's language~\cite{Priestley1965Evolutionary}, the foreground is a \emph{uniformly modulated process} as the entire spectrum shifts in time as a whole.
A natural family of oscillatory functions for the foreground would be $\left\{\sqrt{g_c(t)} e^{2\pi i ft}\right\}$ with $\theta(t,f) = 0$ (or with any time-independent phase $\theta(t,f)$).

We build the simulation from the LDC2a dataset.
First, we compute the SNR with Eq.~\eqref{eq:whittle_snr} at the true parameters for all Galactic binaries and with only the instrumental PSD.
Second, we compute the excess power from sources below a threshold of $\mathrm{SNR} < 7$ and add it to the instrumental PSD.
Third, we recompute the SNR of the superthreshold signals with the new noise that adds the foreground to the instrumental noise.
Fourth, sources that dropped below the SNR threshold are added to the unresolvable set and the total noise curve is adjusted.
We repeat the above steps, achieving convergence after 3-4 iterations.
We then fit the total power of the unresolvable binaries between $0.3$ and $4\,\mathrm{mHz}$ with Eq.~\eqref{eq:foreground_fit} with an inference step using priors from Ref.~\cite{Littenberg:2023xpl} and keep the maximum \textit{a posteriori} values.
A final maximum-likelihood fit with Eq.~\eqref{eq:foreground_modulation} returns the modulation coefficients.
The fitted PSD parameters for different values of $T_{\rm obs}$ are given in Table~\ref{tab:foreground-coeffs}.

\begin{table}[!tbp]
    \centering
    \begin{tabular}{l|c|l|c}
        Parameter & Simulation & \multicolumn{2}{l}{Prior} \\\hline
        \multicolumn{4}{l}{\textit{Signal}}\\
        Initial frequency $f_0$ [mHz]                 & $1$               & \multicolumn{2}{l}{U[$f_{\mathrm{min}}$, $f_{\mathrm{max}}$]} \\
        Frequency derivative $\dot{f}_0$ [Hz\,s$^{-1}$] & $1\times10^{-18}$ & \multicolumn{2}{l}{U[$\dot f_{\mathrm{min}}$, $\dot f_{\mathrm{max}}$]} \\
        Amplitude $\ln A$                             & $-48.68$          & \multicolumn{2}{l}{U[$-54$, $-46$]}\\
        Initial phase $\phi_0$ [rad]                  & $0.5$             & \multicolumn{2}{l}{U[$0$, $2\pi$]} \\
        Ecliptic colatitude $\cos \theta$             & $-0.2083$         & \multicolumn{2}{l}{U[$-1$, $1$]} \\
        Ecliptic longitude $\lambda$ [rad]            & $0$               & \multicolumn{2}{l}{U[$0$, $2\pi$]} \\
        Polarization angle $\psi$ [rad]               & $0.7$             & \multicolumn{2}{l}{U[$0$, $\pi$]} \\
        Inclination $\cos\iota$                       & $0.3$             & \multicolumn{2}{l}{U[$-1$, $1$]} \\\hline
        \multicolumn{4}{l}{\textit{Noise}}\\
        Modulation coefficients {$a_0,a_1,b_1,a_2,b_2$}  & N/A  & \multicolumn{2}{l}{U[$-X$, $X$]} \\\hline
        \multicolumn{4}{l}{\textit{Prior bounds per observation duration $T_{\mathrm{obs}}$}}\\
        Duration $T_{\mathrm{obs}}$ & $f_{\mathrm{min}}$, $f_{\mathrm{max}}$ [mHz] & $\dot f_{\mathrm{min}}$, $\dot f_{\mathrm{max}}$ [$10^{-18}$\,Hz\,s$^{-1}$] & $X$ \\
        $3$ months & {$0.99696$, $1.01205$} & {$-0.493$, $8.36$} & $122$ \\
        $6$ months & {$0.99309$, $1.00691$} & {$-0.485$, $8.20$} & $21.4$ \\
        $1$ year   & {$0.99172$, $1.00596$} & {$-0.482$, $8.18$} & $8$ \\
    \end{tabular}
    \caption{
    Parameters for the simulated LISA data containing a single resolved signal in foreground noise. 
    We list the true simulated values for the signal and the uniform sampling prior for the signal and noise parameters, with lower and upper bounds in square brackets. 
    The five coefficients of the power modulation share a prior.
    The signal frequency bounds correspond to the WDM band containing the signal.
    The WDM likelihood uses all coefficients for both channels, while the stationary frequency-domain one keeps only the constant term $a_{c,0}$.}
    \label{tab:gb_params}
\end{table}

To the foreground noise from unresolved signals we add a single resolvable binary signal, with parameters shown in Table~\ref{tab:gb_params}.
The signal is placed at $1\,\mathrm{mHz}$, where the foreground is a factor of 8 higher than the instrumental noise and varies by a factor of ${\sim} 3{-}4$ over the year.
The remaining signal parameters are chosen such that the signal is anti-correlated with the foreground: when the noise is quiet the signal is loud, and vice versa.
We repeat this process for observations of 3 months, 6 months, and 1 year, when the signal has SNR $\rho_\mathrm{exact} = 12.25, 14.52, 20.26$ respectively.\footnote{The SNR does not accumulate as $\sqrt{T_\mathrm{obs}}$ because of the anti-correlation and noise nonstationarity.}

As the signal is compact in frequency and the noise is defined locally, the analysis is restricted to a narrow frequency range around the signal, with settings provided in Table~\ref{tab:LISA-settings}.
The WDM tiling is based on the optimal calculation of Eq.~\eqref{eq:rms_optimal_tiling}.
Specifically, we consider the $0.98$–$1.02$~mHz band (chosen to be wide enough to cover the bandwidth of interest), fit the spectrum of Eq.~\eqref{eq:foreground_fit} to the data and then evaluate Eq.~\eqref{eq:rms_optimal_tiling} with the logarithmic spectrum derivatives averaged over the band and the full observation time.
Based on the optimal tiling, we adopt the nearest valid tiling that divides the total number of data points.
This adjustment raises the loss by $\lesssim0.2\%$.
All inference models the signal with the model and priors of Table~\ref{tab:gb_params}.
We hold the instrument noise and the time-independent component of the foreground noise fixed to the values given in Table~\ref{tab:foreground-coeffs}.
In practice, the foreground noise is approximately flat across the analysis band.
The time variation of the noise is encoded in $g_c(t)$ whose parameters for each channel are inferred.
The WDM likelihoods contain all five coefficients for both channels, while the frequency-domain one fixes them to zero other than the constant $a_{c,0}$ for each channel.
The wraparound mitigation settings are given in Table~\ref{tab:LISA-settings}.

\begin{table}[!tbp]
    \centering
    \begin{tabular}{l|c|c|c}
        Setting & \multicolumn{3}{c}{Value} \\\hline
        \multicolumn{4}{l}{\textit{Data}}\\
        Observation time $T_{\mathrm{obs}}$ [s]        & $7.884\times10^{6}$ (3 months) & $1.5768\times10^{7}$ (6 months) & $3.1536\times10^{7}$ (1 year) \\
        Cadence $\Delta t$ [s]                         & $5$                 & $5$                  & $5$                  \\
        Sampling frequency $f_s$ [Hz]                  & $0.2$               & $0.2$                & $0.2$                \\
        Nyquist frequency $f_s/2$ [Hz]                 & $0.1$               & $0.1$                & $0.1$                \\
        Samples $N=T_{\mathrm{obs}}/\Delta t$          & $1\,576\,800$       & $3\,153\,600$        & $6\,307\,200$        \\
        Frequency resolution $1/T_{\mathrm{obs}}$ [nHz]& $126.8$             & $63.42$              & $31.71$              \\\hline
        \multicolumn{4}{l}{\textit{WDM grid}}\\
        Time cells $N_t$                               & {$240$} & {$438$} & {$900$} \\
        Frequency cells $N_f$                          & {$6\,570$} & {$7\,200$} & {$7\,008$} \\
        Cell duration $\Delta T=T_{\mathrm{obs}}/N_t$ [d]& {$0.3802$} & {$0.4167$} & {$0.4056$} \\
        Cell bandwidth $\Delta F=1/(2\Delta T)$ [$\mu$Hz]& {$15.22$} & {$13.89$} & {$14.27$} \\\hline
        \multicolumn{4}{l}{\textit{Analysis window }}\\
        WDM frequency bands & $3$ & $3$ & $3$\\
        Lowest frequency [mHz]                         & {$0.98174$} & {$0.97920$} & {$0.97745$} \\
        Highest frequency [mHz]                        & {$1.02727$} & {$1.02080$} & {$1.02023$} \\
        Fourier bins, {$3\,N_t/2$}                 & {$360$} & {$657$} & {$1\,350$} \\
        WDM cells, {$3\,N_t$}                      & {$720$} & {$1\,314$} & {$2\,700$} \\\hline
        \multicolumn{4}{l}{\textit{Wraparound mitigation}}\\
        Edge cut: time bins removed, start / end      & {$5$ / $4$} & {$5$ / $4$} & $0$ / $0$ \\
        Edge cut: WDM cells                            & {$693$} & {$1\,287$} & {$2\,700$} \\
        Exact edges: time bins treated exactly, start / end & $5$ / $4$ & $5$ / $4$ & $0$ / $0$ \\
        Padding: time columns of zeros added           & $16$ & $16$ & $16$ \\
    \end{tabular}
    \caption{Data and analysis settings for the LISA simulation with three observation durations covering a quarter, a half, and a full cycle of the annual foreground modulation.
    The WDM grid shape is set to the optimal tiling from Eq.~\eqref{eq:rms_optimal_tiling}, with the spectrum derivatives averaged over the band of interest.
    The analysis window is restricted to the WDM band containing the signal along with the ones above and below it.
    As $T_\mathrm{obs}$ and $N_t$ grow, the number of Fourier bins in the analysis band, $3 N_t/2$, grows.
    The lowest and highest frequency analyzed are the centers of the first and last Fourier bin, spanning $(3N_t/2-1)/T_\mathrm{obs}$.
    Wraparound mitigation affects a few time-edge bins as indicated.
    No mitigation is required for $T_\mathrm{obs}=1$\,yr as the data are genuinely periodic over that time.
    }
    \label{tab:LISA-settings}
\end{table}

\subsection{Numerical Implementation}
\label{sec:implementation}

We sample the joint posteriors for the signal and noise parameters with \texttt{impulse\_mcmc}~\cite{Johnson2025-fn}, using eight temperatures in a parallel-tempered Markov chain Monte Carlo.
The sampler's default proposals include adaptive Metropolis and differential-evolution.
For the LISA analysis, we further use targeted proposals: a move between phase and polarization, a frequency modulation shift, and a move based on the Fisher matrix eigenvectors~\cite{Littenberg:2020bxy}.
We track the effective sample size and obtain $\gtrsim$3\,000 independent samples on all parameters.
We confirm that the chain is stationary with the split-$\hat{R}$ statistic, which splits it into four segments and checks the between-chain and within-chain statistics~\cite{GelmanRubin1992, Vehtari2021}.
While neither check is sufficient for convergence, they are necessary properties of converged chains.

\paragraph{Zero padding.}
When zero-padding, we extend the grid by $p$ WDM time columns by adding $p N_f/2$ zero samples to each end of the time-domain data.
The values of $\Delta T$ and $\Delta F$ along with the times of the columns holding the data are unchanged. 
We transform the data and the template on the extended grid and restrict to $N_t$ columns.
The covariance $\boldsymbol{\Lambda}$ is evaluated on the extended grid, which is periodic over $T_\mathrm{obs} + p \Delta T$ and thus no longer couples the original data edges.

We account for the zeros that have been added to the data based on the covariance of the padded data.
On the extended grid, the padded data are $\chi(t_j)x(t_j)$ with $\chi = 1$ on the data and $\chi = 0$ on the padding.
The new data WDM coefficients are
\begin{equation}
    w'_{mn} = \sum_j W_{(mn)j}\chi(t_j)x(t_j)\,,
\end{equation}
where $\mathbf{W}$ is the WDM transformation matrix from Eq.~\eqref{eq:cov-transform}.
Their covariance is
\begin{equation}
    \left\langle w'_{mn} w'_{m'n'}\right\rangle = \sum_{jj'}W_{(mn)j}\chi_jC_{jj'}\chi_{j'}W_{(m'n')j'}\,.
\end{equation}
Compared to the original covariance $\boldsymbol{\Lambda}=\mathbf{WCW}^T$, this 
is $\mathbf{W}\boldsymbol{\chi}\mathbf{C}\boldsymbol{\chi}\mathbf{W}^T$ with $\boldsymbol{\chi} = \mathrm{diag}(\chi_j)$.
Introducing $\mathbf{W}^T\mathbf{W}/\Delta t = \mathbf{I}$ on either side of $\mathbf{C}$, we get
\begin{equation}
    \left\langle \mathbf{w'}\mathbf{w'}^T\right\rangle = \frac{1}{(\Delta t)^2}\left(\mathbf{W}\boldsymbol{\chi}\mathbf{W}^T\right)\left(\mathbf{WCW}^T\right)\left(\mathbf{W}\boldsymbol{\chi}\mathbf{W}^T\right) = \mathbf{M} \boldsymbol{\Lambda} \mathbf{M}\,,
\end{equation}
where $\mathbf{M}\equiv\mathbf{W}\boldsymbol{\chi}\mathbf{W}^T/\Delta t$ is the data mask projected onto the WDM grid.
The mask differs from the identity only near the time edges, where its diagonal is the fraction of each WDM basis function $g_{mn}$ whose squared norm is on the data and its off-diagonal entries are the overlaps of neighboring time-frequency cells.
In practice, we compute the covariance matrix $\boldsymbol{\Lambda}$ on the extended grid and compute the likelihood with $\mathbf{M}\boldsymbol{\Lambda}\mathbf{M}$ on the $N_t$ original data columns.
Since $\mathbf{M}$ couples neighboring cells, the final covariance is non-diagonal.

\paragraph{Sinusoid in drifting powerlaw.}
We use Claude Fable 5 to implement the model in \texttt{JAX}~\cite{jax2018github}, vectorize the tempered chains, and evaluate them simultaneously across a GPU.
Noise is simulated with the exact covariance of Eq.~\eqref{eq:wdm_generic_covariance} assuming a time-independent $\theta(t,f)$.
Evaluating that exact covariance during inference is slow; we therefore speed up evaluations by computing it with inverse fast Fourier transforms (FFT).
For time-independent $\theta(t, f)$ the phases cancel, and from Eq.~\eqref{eq:V_matrix} the time-domain covariance matrix is
\begin{equation}
C_{ij} = \Delta f \sum_{k=0}^{N/2}\eta_k \sqrt{S(t_i,f_k)S(t_j,f_k)}\cos\left(2\pi f_k(t_i-t_j)\right)\,.
\end{equation}
For powerlaw noise, Eq.~\eqref{eq:sine_example_psd} implies 
\begin{equation}
    \sqrt{S(t_i,f_k)S(t_j,f_k)} = A_n \left[\frac{\max(f_k,f_b)}{f_\mathrm{ref}}\right]^{\frac{1}{2}(\alpha(t_i) + \alpha(t_j))}= A_n \left[\frac{\max(f_k,f_b)}{f_\mathrm{ref}}\right]^{\alpha(\bar{t}_{ij})}=S(\bar{t}_{ij}, f_k)\,,
\end{equation}
with the average time $\bar{t}_{ij} = (t_i + t_j)/2$.
Thus
\begin{equation}
    C_{ij} = \Delta f \sum_{k=0}^{N/2}\eta_k S\left(\frac{s\Delta t}{2}, k\Delta f\right)\cos\left(\frac{2\pi k \ell}{N}\right)= \frac{N\Delta f}{2}\mathrm{IRFFT}\left[S\left(\frac{s\Delta t}{2}, k\Delta f\right)\right]\,,
\end{equation}
where we have introduced the index $s = i + j \in [0, 2N-2]$ and the lag index $\ell = |i-j|\in [0, N-1]$.
In the second equality, we have written the covariance as an inverse real FFT in the \texttt{numpy} convention.
For each value of the midpoint $s$, a single IRFFT gives the entire set of lags $\ell$.
Since the order of complexity for the IRFFT is $\mathcal{O}(N\log N)$, the overall complexity is $\mathcal{O}(N^2\log N)$ for the $2N-1$ midpoints.
We then transform the time-domain covariance to the WDM domain.

\paragraph{Galactic binary in cyclostationary foreground noise.}
We use Claude Fable 5 to implement the model in \texttt{JAX}~\cite{jax2018github}, including the waveform, TDI, and proposal distributions from \texttt{GLASS}~\cite{Littenberg:2023xpl}.
The software re-implementation has been verified against the original, returning matching values across a broad range of input parameters.
We have packaged this as a standalone \texttt{Python} package called \texttt{binary\_galactica}, available at \url{https://github.com/AaronDJohnson/binary_galactica}~\cite{binary_galactica}.
The binary waveform is implemented via the ``TDI on the fly'' method~\cite{Cornish:2025jfu}.

The analysis bandwidth is relatively small; with at most $3$ WDM frequency bands and $900$ time bins, Table~\ref{tab:LISA-settings}, the full covariance of Eq.~\eqref{eq:wdm_generic_covariance} can be computed with a GPU in a reasonable time.
In comparison to this exact likelihood, the cost of the tri-diagonal and diagonal likelihoods is small even when the edge-time columns are treated exactly.
Implementations of all likelihoods and covariances, together with the scripts that produce our results, are available at 
\url{https://github.com/AaronDJohnson/wdm_models_scripts}~\cite{wdm_models_scripts}.

\section{Results}
\label{sec:results}

In this section we first explore the covariance structure of the example noise models in Sec.~\ref{sec:covariance_structure}, then analytically assess the diagonal covariance approximation in Sec.~\ref{sec:result-tiling}, and then present inference results for the sinusoidal signal, Sec.~\ref{sec:sinusoid_signal} and a LISA Galactic binary, Sec.~\ref{sec:galactic_binary}, under different assumptions about the noise covariance.

\subsection{Covariance structure}
\label{sec:covariance_structure}

\begin{figure}[!tbp]
    \centering
    \includegraphics[width=\linewidth]{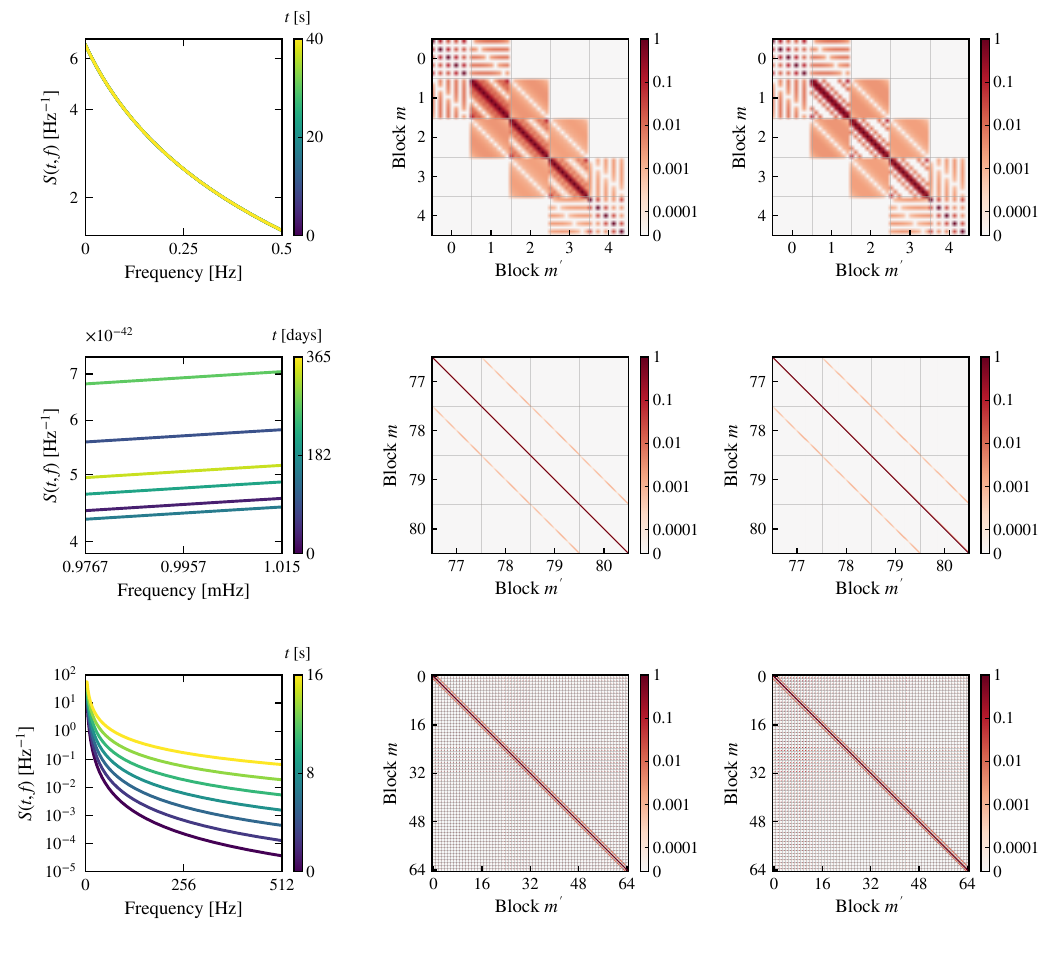}
    \caption{Noise spectra (left), normalized covariance $\boldsymbol{\rho}$ (middle), and normalized inverse covariance $\boldsymbol{Q}$ (right).
    From top to bottom, we have stationary noise, cyclostationary noise in LISA's A channel for a Galactic-binary foreground, and drifting powerlaw noise.
    The left column shows $S(t, f)$ at times indicated by the colorbar.
    The middle and right columns show the absolute value of $\boldsymbol{\rho}$ and $\mathbf{Q}$ on a logarithmic scale, with zero set to white.
    Axis ticks give the frequency index of rows $m$ and columns $m'$.
    Each frequency block contains $N_t \times N_t$ time entries, separated by thin lines.
     For stationary noise we show the full $50\times 50$ unpacked grid including vanishing edge bins.
    For LISA noise we show a $3\,200\times 3\,200$ sub-block of the covariance, covering four frequency bands around 1\,mHz, corresponding to $m \in [77, 80]$, where the foreground noise varies the most.
    Each frequency block contains $800\times 800$ time cells.
    For the drifting powerlaw, we display the full $16\,640\times 16\,640$ matrix with 65 frequency blocks and 256 time cells.
    For stationary noise, both the covariance and its inverse are block tri-diagonal.
    Cyclostationary LISA noise remains approximately block tri-diagonal due to weak time and frequency dependence.
    For the drifting powerlaw, correlations between distant frequency bins are concentrated around the time-edge bins.}
    \label{fig:covariance_examples_log}
\end{figure}

To gain some intuition, we begin by inspecting the covariance of the drifting powerlaw and LISA foreground noise. 
We focus on the time-dependent noise and omit the signals and the instrument PSD in LISA's case.
For comparison, we also consider a stationary Gaussian process with
\begin{equation}
    S(f) = \frac{1}{f + f_c}\,,
\end{equation}
where $f_c = 0.15\,\mathrm{Hz}$.
We use $T_\mathrm{obs} = 40\,\mathrm{s}$ of data sampled at $\Delta t = 1\,\mathrm{s}$, for $N = 40$ samples.
The WDM grid has $(N_f, N_t) = (4, 10)$, with $\Delta T = 4\,\mathrm{s}$ and $\Delta F = 1/8\,\mathrm{Hz}$.
The covariance is computed exactly with Eq.~\eqref{eq:wdm_generic_covariance} and no timescale assumptions.
In the LISA case, we select the $T_\mathrm{obs} = 365\,\mathrm{d}$ simulation from Tables~\ref{tab:LISA-settings} and~\ref{tab:foreground-coeffs} and plot a narrow band around $1\,\mathrm{mHz}$, where the foreground dominates the total noise and the spectrum is approximately flat.

The generic WDM covariance is real, symmetric, and manifestly positive semi-definite by its Gramian structure.
The vanishing edge-band coefficients would make the matrix singular~\cite{Pearson:2025wfd,Vajpeyi:2026msr} unless we pack the nonzero Nyquist elements into the zero DC elements.
Below, we show the unpacked matrices, but use an intermediate packing to invert them.
To make structural comparisons, we plot the normalized correlation matrix of Eq.~\eqref{eq:correlation_matrix} and the similarly-normalized precision matrix $\mathbf{\Lambda}^{-1}$, denoted as $\mathbf{Q}$.
Both the correlation matrix, $\boldsymbol{\rho}$, and $\mathbf{Q}$ have a unit diagonal and since they are positive definite their entries fall within $[-1, 1]$.
Figure~\ref{fig:covariance_examples_log} displays the spectra and selected matrix entries.
Within a diagonal frequency block ($m=m'$), off-diagonal entries couple different times, while off-diagonal frequency blocks ($m\neq m'$) couple different bands.

For the stationary noise we display the full $50\times 50$ unpacked matrices,\footnote{The flattened-index covariance matrix has dimension $(N+N_t)\times(N+N_t)$.} with inactive edge-band slots, $m=0$ and $m=N_f$, shown as zero.
As the frequency-domain covariance is diagonal, the WDM-domain covariance is limited to a block tri-diagonal structure with $|m-m'|\le 1$~\cite{Cornish:2025awt, Johnson:2026rrn}.
Inverting such a matrix leads to the same structure, as can immediately be seen by first inverting the diagonal Fourier-domain matrix and then transforming to the WDM domain~\cite{Johnson:2026rrn}.
Indeed, entries with $|m-m'|>1$ vanish to machine precision.
For this example, same-frequency bands ($m=m'$) reach up to $9.30\%$ of the diagonal ($m=m'$ and $n=n'$) entries in $\boldsymbol{\rho}$ and $9.19\%$ in $\boldsymbol{Q}$.
Adjacent band ($m\neq m'$) entries reach $0.67\%$ and $0.61\%$ respectively.
These numbers depend on the PSD spectral slope and the spectrum variation across the Meyer window, with a flatter spectrum having smaller off-diagonal terms.

For LISA cyclostationary noise we show a $3\,200\times3\,200$ sub-block of the covariance: 
the full $N_t=800$ time cells in four frequency bands near 1\,mHz.
In this case, same-band ($m=m'$) time correlations reach $0.29\%$ of the diagonal in both $\boldsymbol{\rho}$ and $\mathbf{Q}$, while adjacent-band ($m\neq m'$) entries reach maxima of $<0.1\%$ of the diagonal.
The slow time variation of the foreground and the weak spectral variation across the narrow band considered lead to weak correlations and an approximately diagonal covariance.
Additionally, there are no large correlations due to the wraparound effect discussed in Sec.~\ref{sec:likelihoodmethod} as the noise is periodic over the observation.

For the drifting powerlaw, we show the full $16\,640\times16\,640$ unpacked matrices, with all time cells and frequency bands.
Now same-band ($m=m'$) correlations reach $95\%$ of the diagonal in both $\boldsymbol{\rho}$ and $\mathbf{Q}$, while adjacent-band ($m\neq m'$) correlations reach $73\%$ and $92\%$ respectively.
Furthermore, correlations extend well beyond neighboring bands: entries with $|m-m'|>1$ reach $42\%$ in $\boldsymbol{\rho}$ and $83\%$ in $\mathbf{Q}$.
These large entries appear near the data's time boundary, $n=0$ and $n=N_t-1$, for every frequency band, corresponding to the wraparound effect.
The WDM transformation, like the discrete Fourier transform, is susceptible to wraparound, even after windowing: the time extent of the WDM window wraps around from the end of the data back to the start, correlating them.
The effect is particularly noticeable for this PSD that varies by a factor of $\mathcal{O}(10^2-10^3)$, depending on frequency, over the time of observation.

\subsection{Optimal WDM tiling and diagonal covariance approximation for LISA noise}
\label{sec:result-tiling}

\begin{figure}[!tbp]
    \centering
    \includegraphics[width=0.82\linewidth]{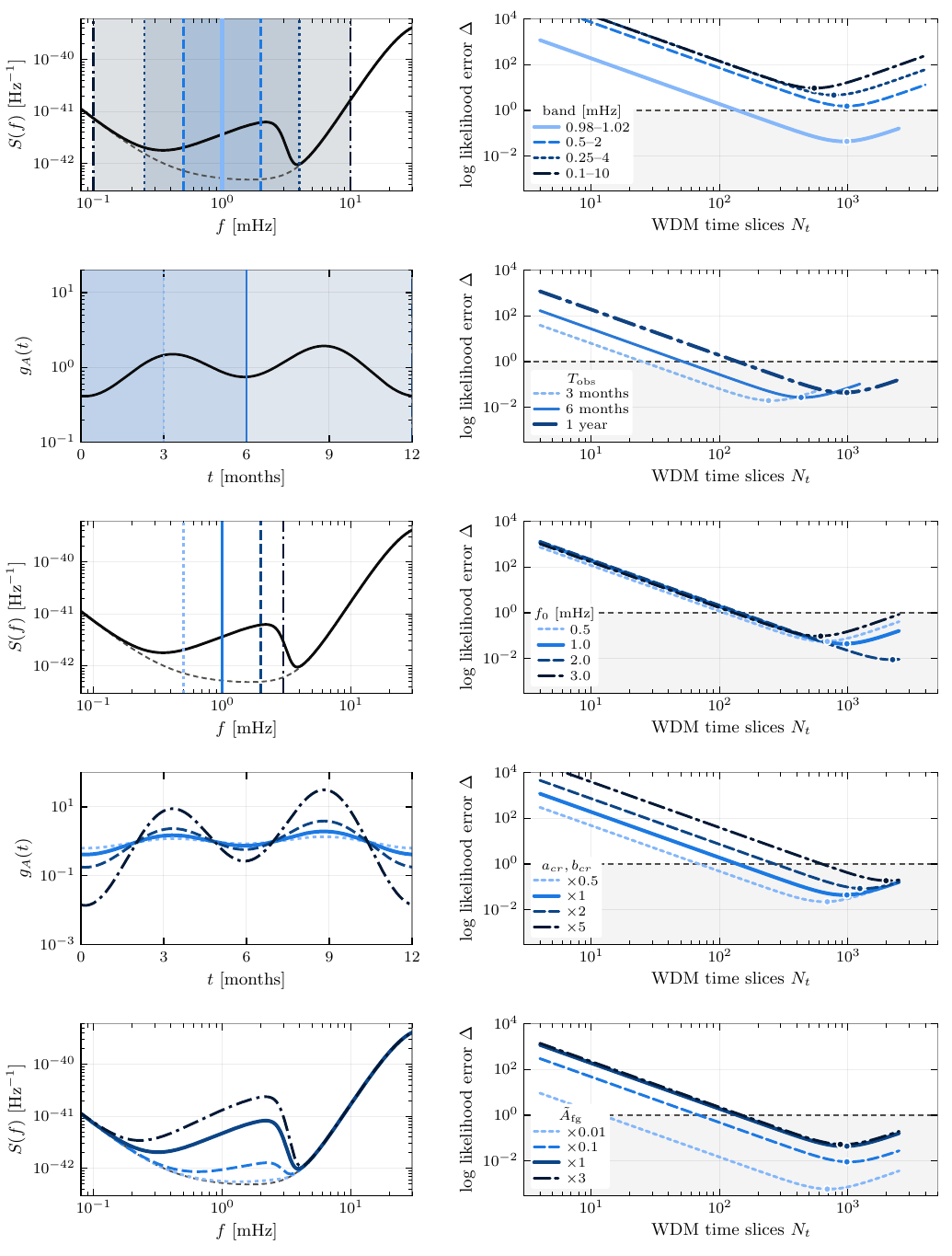}
    \caption{Log likelihood error from approximating the WDM covariance as diagonal, Eq.~\eqref{eq:error-Nt}, for the LISA cyclostationary noise of Sec.~\ref{sec:UCB-methods}. Each row explores a different effect depicted in the left column and the corresponding error as a function of the WDM tiling $N_t$ in the right column.
    From top to bottom we vary: the analysis band, denoted with shading on the spectrum; the observation duration, denoted along the annual modulation; the source frequency, denoted by vertical lines on the spectrum; the modulation depth; and the foreground amplitude.
    Circles on the right mark the minimum of each curve and the grey shading corresponds to the $\Delta<1$ threshold. The curves terminate where the analysis band reduces to a single frequency cell, where the WDM representation is a band-passed time series.}
    \label{fig:loglikeloss_lisa}
\end{figure}

The optimal WDM tiling criterion of Sec.~\ref{sec:optimal} provides a leading-order estimate for the error of approximating the spectrum $S(t,f)$ as smooth across a cell and the covariance as diagonal.
By construction, it does not consider the wraparound correlations.
Here, we explore that error for the simulated noise of Sec.~\ref{sec:methods}.
For the drifting powerlaw of Sec.~\ref{sec:monochoromatic-methods}, the leading-order approximation of Eq.~\eqref{eq:evspectrum-expansion} breaks for any tiling as the frequency- and time-dependence parameters, $\varepsilon_f$ and $\varepsilon_t$, are never both $<0.1$. 
This is in addition to the large wraparound correlations depicted in Sec.~\ref{sec:covariance_structure}.

The LISA cyclostationary noise of Sec.~\ref{sec:UCB-methods} is smooth-enough in frequency and stationary-enough in time that the perturbative expansion holds broadly. 
Figure~\ref{fig:loglikeloss_lisa} shows the log likelihood error (right) as a function of $N_t$ as various noise and analysis settings are varied from top to bottom (depicted on the left).
Unless varied, we consider $T_{\rm obs}=1\,$yr, the $0.98$--$1.02\,$mHz analysis band, the noise parameters of Table~\ref{tab:foreground-coeffs}, and combine the $A$ and $E$ channels.
The logarithmic spectrum derivatives of Eq.~\eqref{eq:error-Nt} are RMS averaged over the analyzed time-frequency region.
Each curve for the likelihood error terminates at the value of $N_t$ for which there is only one WDM frequency band.

\begin{itemize}
\item In the first row we vary the analysis bandwidth, keeping it centered at $1\,$mHz.
At the nominal band of $0.98$--$1.02\,$mHz, the optimal setting is $N_t^\star=982$, consistent with Table~\ref{tab:LISA-settings} once rounded to the nearest divisor of $N$, and the corresponding error is $0.044$, well below the threshold of 1.
The diagonal covariance is therefore an admissible approximation when analyzing such a small band.
As the analysis band grows, the optimal $N_t$ decreases.
This is because the spectrum variation is stronger, and thus the width of the frequency cells $\Delta F$ needs to decrease.
The diagonal approximation is inadmissible for the larger-bandwidth analyses, with the error reaching $1.5$, $4.6$ and $9.3$ for the $0.5$--$2$, $0.25$--$4$ and $0.1$--$10\,$mHz bands respectively.

\item The second row varies the observation duration.
As $T_{\rm obs}$ increases, so does the amount of noise nonstationarity probed by the analysis.
The optimal $N_t$ therefore increases from $241$ at 3 months to $982$ in 1\,yr, nearly in proportion to $T_{\rm obs}$ so that the cell duration stays at ${\simeq}0.4\,$d.
As the cell duration, and thus the amount of noise variation across a time cell, remains constant, the optimal error remains below threshold throughout, growing only from $0.020$ to $0.044$.

\item The third row changes the central frequency of the analysis band across the foreground, $f_0=0.5$, $1$, $2$ and $3\,$mHz.
Both the optimal $N_t$ and the error are a weak function of $f_0$, with the exception of $2\,$mHz.
This corresponds to the foreground maximum, where $\partial_f S$ nearly vanishes and the optimum moves to larger $N_t$ and a lower error.

\item The fourth row varies the modulation depth by scaling the coefficients $a_{c,r},b_{c,r}$ of Eq.~\eqref{eq:foreground_modulation} at fixed $a_{c,0}$.
A larger modulation requires a larger $N_t$ as the noise is more nonstationary, from $N_t^\star=691$ to $1981$.
However even with a $5\times$ larger modulation, the error remains below threshold, growing only from $0.022$ to $0.19$.

\item The fifth row varies the foreground amplitude by scaling $\tilde{A}_{\rm fg}$ in Eq.~\eqref{eq:foreground_fit}.
A stronger foreground has a less smooth spectrum and thus the optimum error increases from $5.9\times10^{-4}$ to $0.053$.
The optimal $N_t$, however, remains constant near $10^3$.
This is because the optimal $N_t$ balances the time and frequency spectrum derivatives.
Both grow with a stronger foreground so that their ratio is nearly independent of $\tilde{A}_{\rm fg}$.
Even with a $3\times$ stronger foreground, the error is below threshold.
\end{itemize}

Overall, a diagonal WDM covariance is admissible for narrow-band analyses, such as a resolvable Galactic binary, but will not be sufficient for full-bandwidth analyses.

\subsection{Sinusoid signal in drifting noise}
\label{sec:sinusoid_signal}

\begin{figure}[!tbp]
    \centering
    \includegraphics[width=\columnwidth]{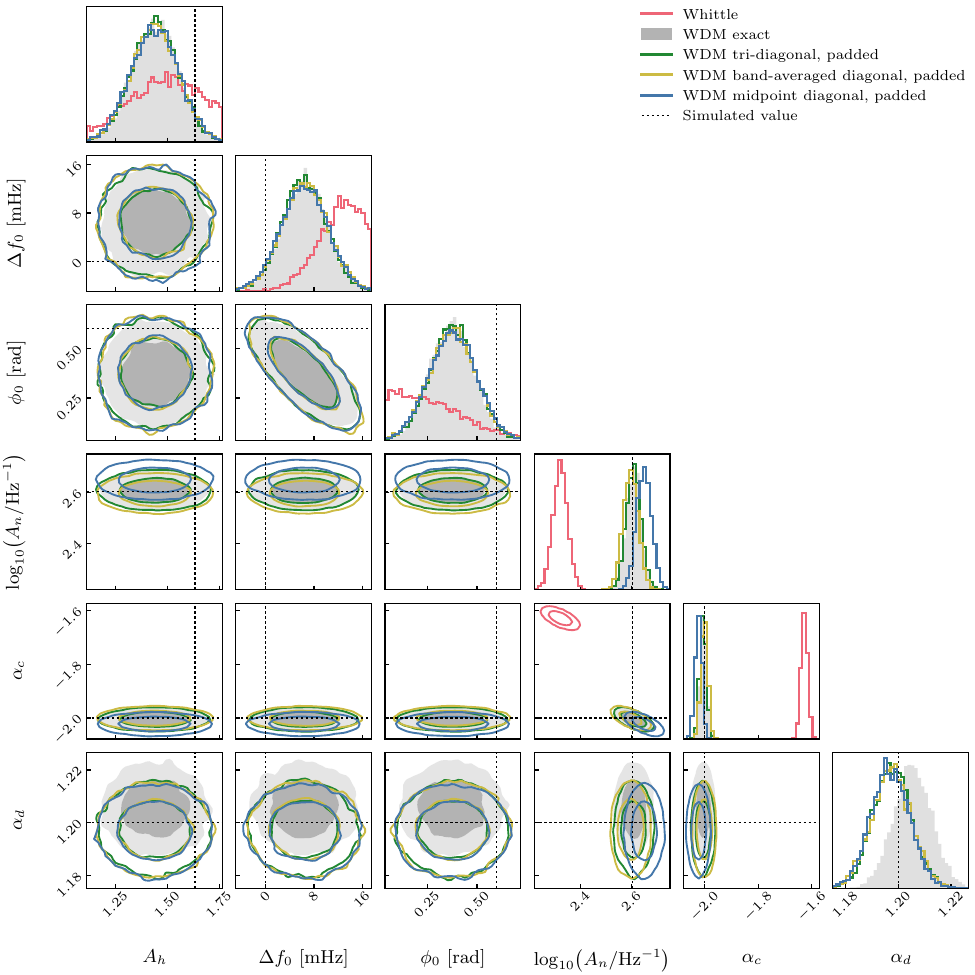}
    \caption{Marginalized posteriors for the signal and noise parameters, defined in Table~\ref{tab:sine_params}, of a $33\,$Hz sinusoid signal in drifting powerlaw noise, Sec.~\ref{sec:monochoromatic-methods}.
    We model the data with different likelihoods from Table~\ref{tab:likelihood-matrix} that probe the impact of covariance approximations.
    We show the WDM likelihood with the exact covariance in gray filled, the frequency-domain stationary Whittle likelihood in red, and the zero-padded WDM likelihoods with the tri-diagonal, band-averaged diagonal, and midpoint diagonal covariances in green, yellow, and blue respectively.
    The Whittle likelihood includes a constant powerlaw index, which we plot on the $\alpha_c$ panel.
    Dashed lines denote the simulated values and $\Delta f_0$ is the difference from the simulated frequency.
    Posteriors are offset from the simulated values due to the Gaussian noise realization.}
    \label{fig:sine_domain}
\end{figure}

Following the methods of Sec.~\ref{sec:monochoromatic-methods}, we simulate and infer sinusoid signals, covering low, mid, and high frequencies, in drifting powerlaw noise.
Per Sec.~\ref{sec:covariance_structure}, we expect strong covariance correlations, reaching 95\% within a frequency band, and correlated time edges across distant frequency bands.
Indeed the strong noise nonstationarity costs the stationary Whittle likelihood a significant amount of SNR.
While all signals are simulated with $\rho_\mathrm{exact} = 15$, the stationary assumption amounts to $\rho_{\mathrm{whittle}}=7.8$ for the 33\,Hz signal, further reduced to $4.6$ and $3.0$ for the 137\,Hz and 385\,Hz signals respectively.
The Whittle-likelihood analysis fails to detect the latter two signals, while yielding similar noise posteriors to the 33\,Hz signal simulation. 
In what follows, we therefore restrict to the 33\,Hz signal.

Posteriors for the signal and noise parameters are shown in Fig.~\ref{fig:sine_domain}.
While the Whittle-likelihood analysis recovers the signal (amplitude constrained away from zero), it does so at reduced SNR leading to broader and biased posteriors.
The noise inference is also heavily biased, both in amplitude and slope.
In contrast, all WDM likelihoods recover the signal with essentially identical posteriors, shown here with padding time-edge mitigation.
Regarding noise inference, no padded approximate WDM likelihood exactly matches the exact posterior obtained with the full covariance, though all posteriors are broadly consistent.
The powerlaw index time derivative, $\alpha_d$, is slightly underestimated, though as we show in Fig.~\ref{fig:sine_edge} below, for the tri-diagonal and band-averaged diagonal likelihood this is due to the zero-padding and not the covariance approximation.
In addition, the midpoint diagonal likelihood's posteriors are offset in the noise amplitude, $\log_{10}A_n$, and powerlaw index intercept, $\alpha_c$.
The midpoint likelihood offsets are consistently recovered for every edge treatment, showing that the midpoint approximation remains poor due to its approximate diagonal terms. 

Figure~\ref{fig:sine_edge} compares the time-edge treatments for the tri-diagonal covariance, with similar results obtained for the band-averaged diagonal case.
Without mitigating the large time-edge correlations, both signal and noise parameters are heavily biased and inconsistent with their true values (the noise parameter posteriors lie beyond the plot scale).
Treating the edge columns exactly, instead, leads to posteriors matching the full-covariance analysis.
However, not only is this edge-handling strategy the most expensive, it is also generally not available for real data as it assumes knowledge of the time dependence of the oscillatory family phase, $\theta(t, f)$.
Excising the time-edge bins leads to some information and SNR loss: the tri-diagonal likelihood evaluated at the simulated parameters yields $\rho_\mathrm{tri, cut} = 14.25$ compared to the exact value of 15.
Padding the data or excising the edge both return similar posteriors to the exact covariance, with small differences in $\alpha_d$ and $A_h$ respectively.
Overall, time-edge mitigation with the tri-diagonal likelihood is effective and also essential for noise that varies significantly between the beginning and end of data.

\begin{figure}[!tbp]
    \centering
    \includegraphics[width=\columnwidth]{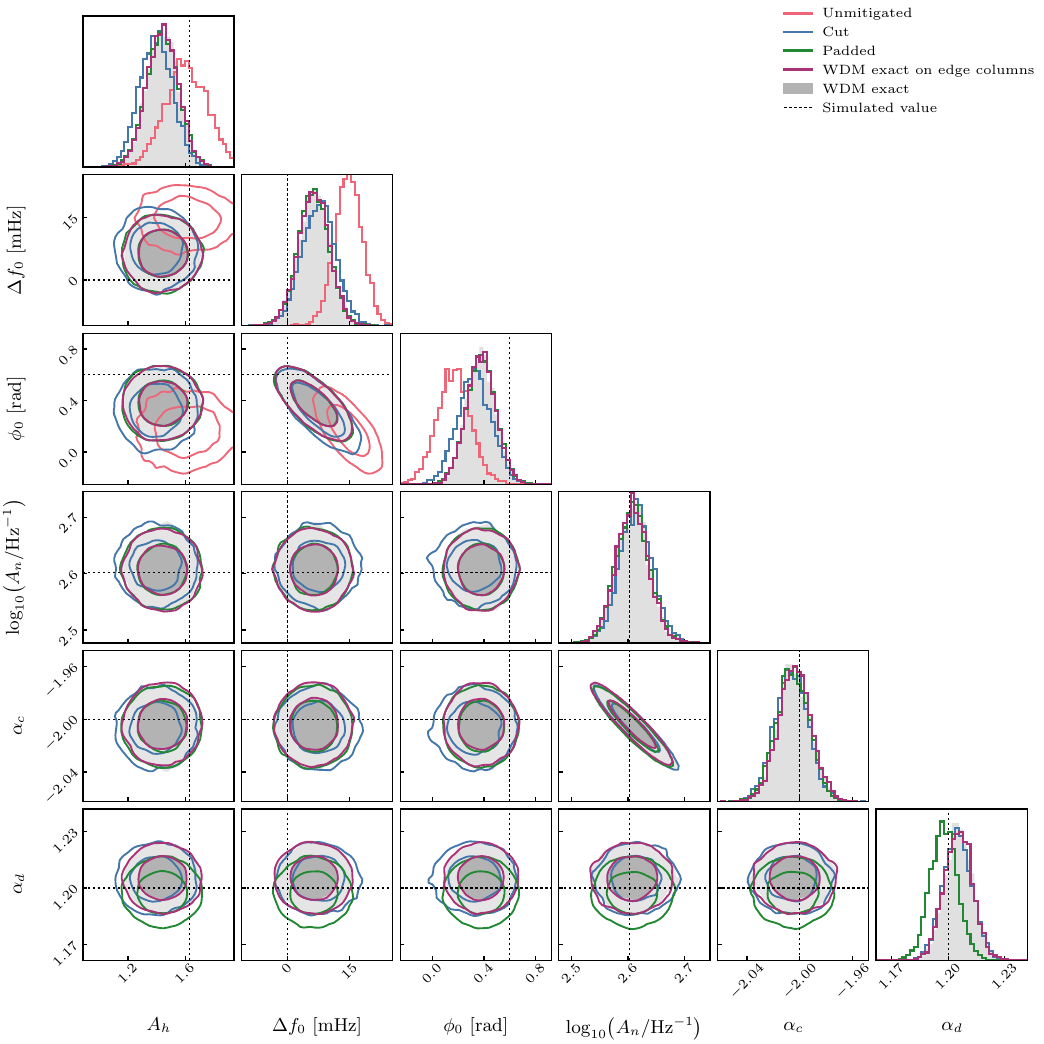}
    \caption{Same as Fig.~\ref{fig:sine_domain}, but with likelihoods from Table~\ref{tab:likelihood-matrix} that use the tri-diagonal covariance with different time-edge correlations mitigation.
    The exact covariance result is shown in gray filled for comparison.
    Leaving the time-edge covariance unmitigated is shown in red, cutting the time edges is shown in blue, padding with zeros is shown in green, and replacing the affected edge bins with the exact covariance is shown in purple. 
    The unmitigated noise posteriors fall beyond the scale of the plot and are thus not shown.
    }
    \label{fig:sine_edge}
\end{figure}

\subsection{Single LISA Galactic binary in cyclostationary noise}
\label{sec:galactic_binary}

Switching to the LISA Galactic binary simulation of Sec.~\ref{sec:UCB-methods}, we investigate the impact of different covariance approximations and observation durations.
Unlike the sinusoid of Sec.~\ref{sec:sinusoid_signal}, the analysis now covers such a small bandwidth that the instrument and foreground noise PSDs are both nearly flat.
With only three WDM bands analyzed, the optimal tiling prioritizes small time bins: 240 to 900 time cells depending on $T_{\rm obs}$ each with duration ${\simeq}0.4\,$d.
Moreover, the noise levels at the start and end of the data differ by ${\cal{O}}(10)$ for both TDI channels compared to $\mathcal{O}(10^2$--$10^3)$ for the powerlaw noise.

\begin{figure}[!tbp]
    \centering
    \includegraphics[width=\columnwidth]{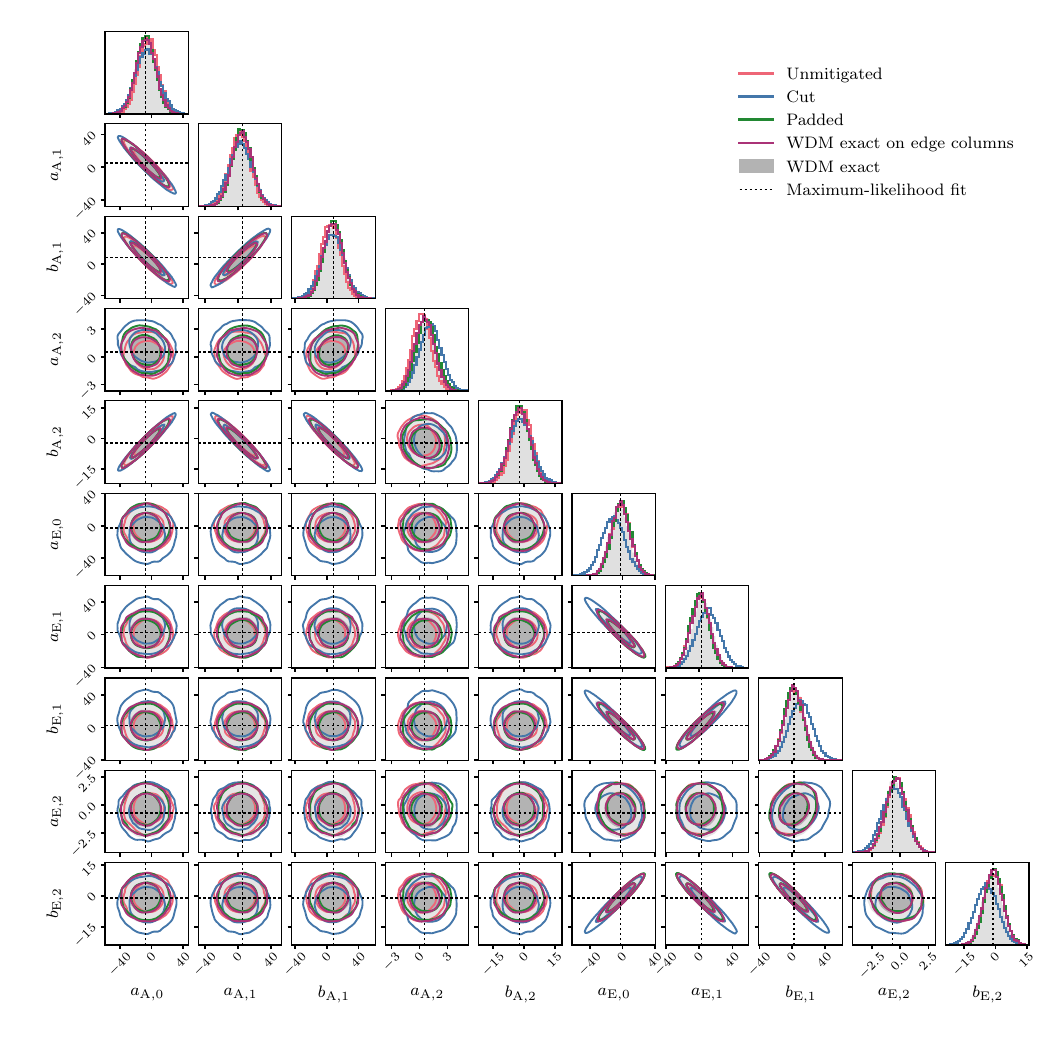}
    \caption{Marginalized posteriors for the noise parameters, defined in Table~\ref{tab:gb_params} and Eq.~\eqref{eq:foreground_modulation}, of the annual modulation $g_c(t)$ of foreground noise with 3\,months of data for the A and E channels.
    We model the data with different likelihoods from Table~\ref{tab:likelihood-matrix} that use the tri-diagonal covariance with different time-edge correlations mitigation.
    The exact covariance result is shown in gray filled for comparison.
    Leaving the time-edge covariance unmitigated is shown in red, cutting the time edges is shown in blue, padding with zeros is shown in green, and replacing the affected edge bins with the exact covariance is shown in purple. }
    \label{fig:gb_3mo_noise_corner}
\end{figure}

Wraparound correlations are therefore expected to be weaker, confirmed in Fig.~\ref{fig:gb_3mo_noise_corner} for the 3-month observation that has the largest noise level change.
We find that all time-edge mitigation strategies return consistent posteriors to the exact-covariance analysis for the noise parameters that specify the modulations $g_c(t)$ for the A and E channels.
For this result we use the tri-diagonal covariance and explore: applying no treatment of the time edges, cutting the affected time columns, zero-padding the front and back of the data, and treating the edges exactly.
Differences are only visible when excising time-edge data where the cut widens the posterior as we remove ${\sim}3.75$\% of the data.
Periodic wraparound of the data correlates cells whose noise levels differ by less than a factor of 10. 
Thus correlations remain small, and even leaving the edges unmitigated is acceptable.
For longer observations, the difference in noise levels is even lower, with the 12-month observation being genuinely periodic.

\begin{figure}[!tbp]
    \centering
    \includegraphics[width=\columnwidth]{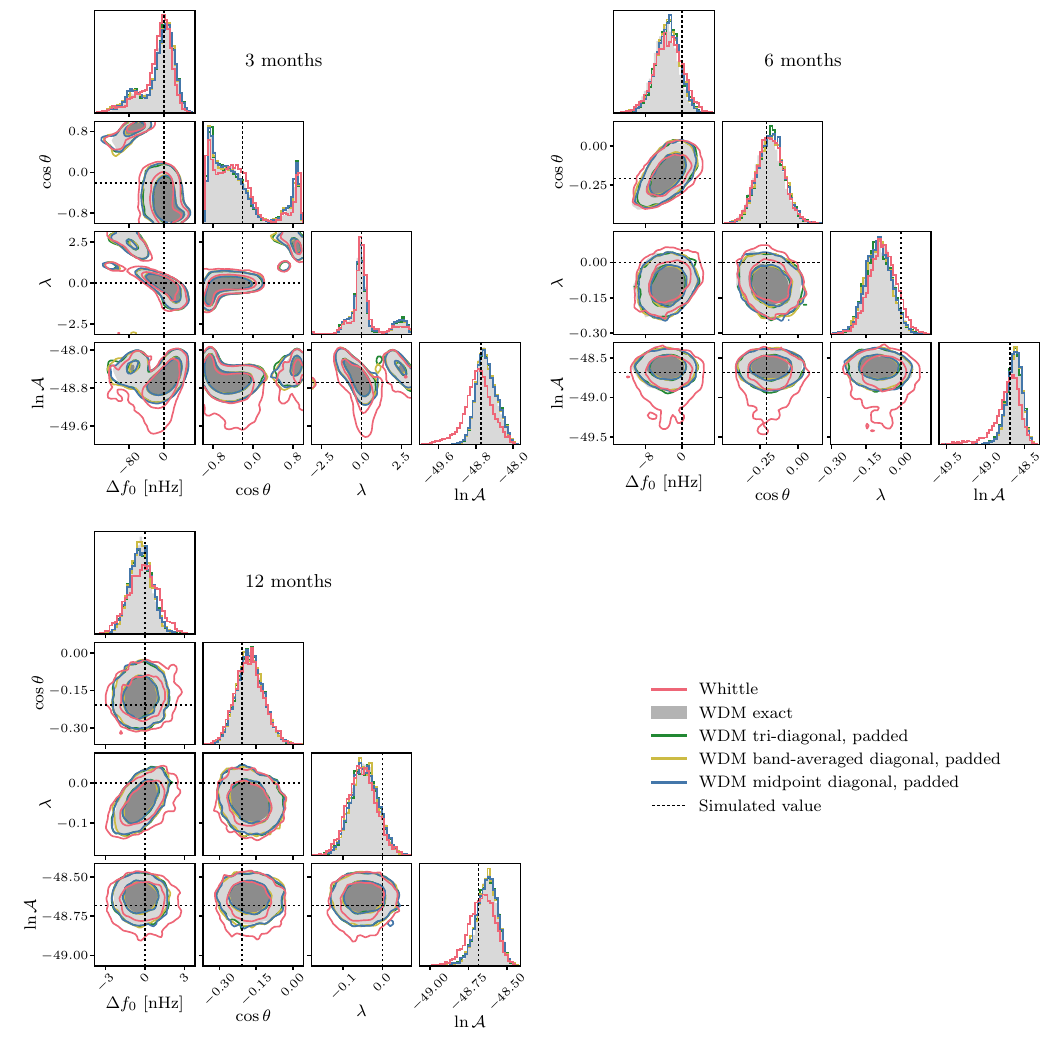}
    \caption{Marginalized posteriors for selected signal parameters, defined in Table~\ref{tab:gb_params}, of a Galactic binary in LISA cyclostationary noise, Sec.~\ref{sec:UCB-methods}, for different observation lengths of $3$, $6$ and $12$ months.
    We model the data with different likelihoods from Table~\ref{tab:likelihood-matrix} that probe the impact of covariance approximations.
    We show the WDM likelihood with the exact covariance in gray filled, the frequency-domain stationary Whittle likelihood in red, and the zero-padded WDM likelihoods with the tri-diagonal, band-averaged diagonal, and midpoint diagonal covariances in green, yellow, and blue respectively.
    Dashed lines denote the simulated values and $\Delta f_0$ is the difference from the simulated frequency.}
    \label{fig:gb_signal_corner}
\end{figure}

In Fig.~\ref{fig:gb_signal_corner}, we show posteriors for a subset of the parameters of the resolved Galactic binary using different covariance approximations.
For all values of $T_{\rm obs}$, the frequency-domain Whittle likelihood, which assumes stationary noise, 
loses SNR and returns broader posterior marginals.
Specifically for the maximum-a-posteriori sample, while the exact-covariance SNR is $\rho_\mathrm{exact,MAP} = 12.64, 14.52, 20.36$, the frequency-domain analysis recovers $\rho_{\mathrm{whittle, MAP}}=9.63, 12.77, 17.95$
for $T_{\rm obs}$ of 3, 6, and 12 months respectively.
The stationary-noise approximation leads to a $12\%-24\%$ SNR reduction.
In contrast, all WDM-domain analyses return SNRs within 1\% to 3\% of the exact value and indistinguishable posteriors.

\begin{figure}[!tbp]
    \centering
    \includegraphics[width=\columnwidth]{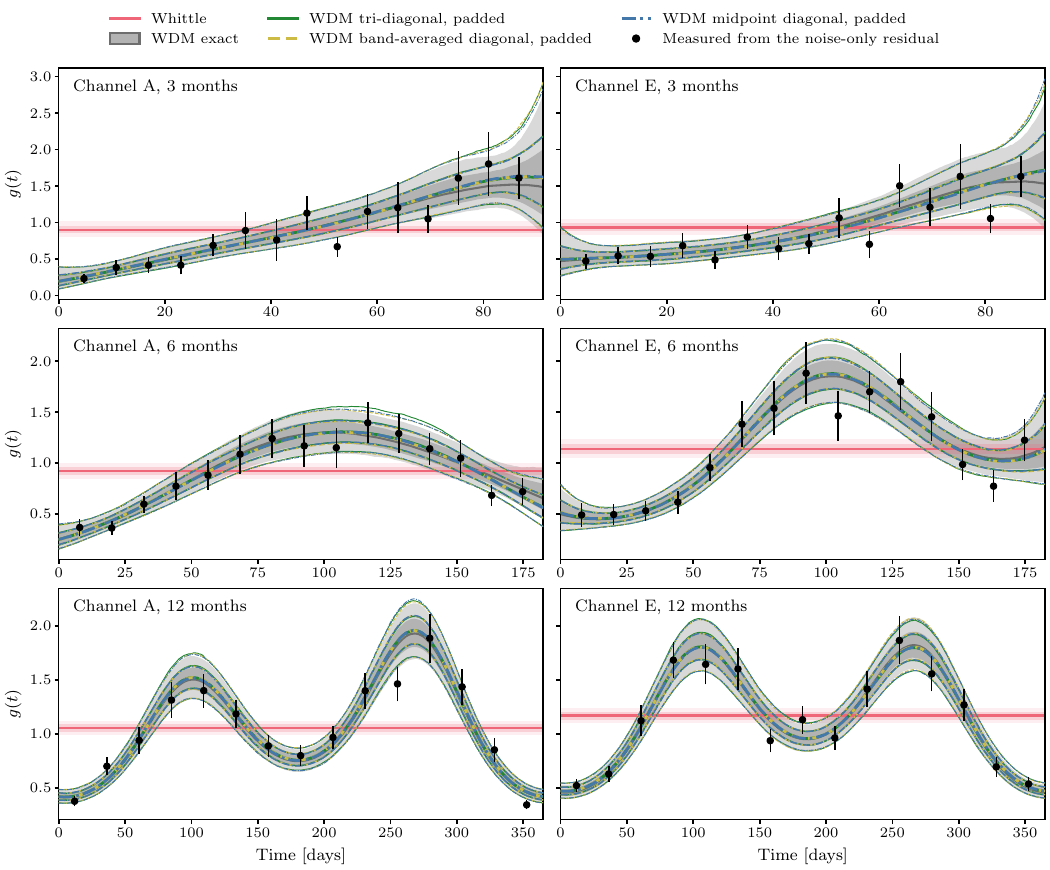}
    \caption{Posteriors for the reconstruction of the foreground modulations from the analyses of Fig.~\ref{fig:gb_signal_corner} for both TDI channels (columns) and observation durations (rows).
    In black we show the measured modulation from the noise-only data over the 3 WDM frequency bands and 15 time divisions (mean and standard error of the mean).
    Lines and contours correspond to the median and the 68\% and 95\% credible levels.
    Due to its assumption of stationarity, the frequency-domain Whittle likelihood recovers the average noise over the period of observation.}
    \label{fig:gb_noise_draws}
\end{figure}

Posterior reconstructions of the time-varying modulation $g_c(t)$ for each TDI channel and observation duration are shown in Fig.~\ref{fig:gb_noise_draws}.
The black points show the noise modulation as measured directly from the data.
To compute it, we transform the noise-only data, i.e., the simulated data without the resolved Galactic binary, to the WDM domain and obtain coefficients $w_{mn}$.
Inverting the cell-center variance of Eq.~\eqref{eq:wdm_variance_cell_center} gives an estimate of the time-variation at that cell's time, $\hat g_{mn}=[2w_{mn}^2-S_\mathrm{inst}(f_m)]/S_\mathrm{fg}(f_m)$.
As a single cell is noisy, we average over the three frequency bands and 15 time blocks, excluding the time edges, and plot the mean $\hat g_b$ and standard error $s_b/\sqrt{n_b}$, where $s_b$ is the scatter of the $n_b$ single-cell estimates in the block,
\begin{equation}
    s_b^2=\frac{1}{n_b-1}\sum_{m,n\in b}\left(\hat g_{mn}-\hat g_b\right)^2\,.
\end{equation}
With 45--180 cells per block, relative errors range from 10\% to 40\%.

All WDM-domain analyses successfully reconstruct the time variation of the data, with essentially indistinguishable posteriors.
Moderate deviations from the exact result are only visible near the time edges of the 3-month observation case.
Constraints also tighten as the observation time increases, especially close to the time edges, as more of the modulation becomes visible.
In the frequency domain, the stationary Whittle analysis is incapable of tracking the time-variation and instead recovers the average noise level.

\section{Conclusion}
\label{sec:conclusions}

Building on the theory of oscillatory processes by \citet{Priestley1965Evolutionary, Priestley1981}, we have developed evolutionary power spectra for nonstationary noise in the WDM basis.
When the nonstationary timescale is long compared to the temporal extent of the filter, $B_\phi\ll B_x$, the evolutionary PSD $S(t, f)$ is a distribution of power over physical frequencies at each time, non-negative by construction, and its band-averaged form can be estimated from a single noise realization.
In the WDM basis, the noise covariance to leading order in $B_\phi / B_x$ is block tri-diagonal, as for stationary noise, with diagonal and off-diagonal elements given by integrals of the spectrum over the Meyer filter, and independent of the choice of oscillatory function family.

This framework provides a rigorous formalism for the evolutionary PSD in terms of a controlled expansion in $B_\phi / B_x$. 
By comparison, the alternative Wigner--Ville dynamic spectrum can take negative values, and positivity must be imposed \textit{post hoc} by smoothing, at the cost of resolution~\cite{Cohen1995}.
In another example, \citet{Cornish:2026tjt} used a Gramian construction, which is an oscillatory process with time-independent $\theta (t,f)$, cf. Sec.~\ref{sec:cov-generation}.
A positive definite covariance and physical-frequency interpretation follow as that construction is a subclass of the oscillatory process framework.
The theory of \citet{Priestley1981} makes it explicit that a positive $S(t, f)$ alone does not determine the covariance, and that only the leading-order covariance in $B_\phi/B_x$ is independent of the choice of oscillatory functions.

The same controlled expansion quantifies the error of approximating the covariance.
The optimal tiling of Sec.~\ref{sec:optimal} balances the cell duration, which hides the nonstationarity, against the bandwidth, which hides the spectral color, and gives the leading-order log likelihood error of a diagonal covariance.
For LISA's cyclostationary Galactic foreground, a diagonal covariance is appropriate for a narrow-band analysis around a resolved binary, with a log likelihood error of $\simeq0.02-0.04$ at the optimal tiling, but not for broad-band analyses where the error exceeds unity.
Coarse-graining the time-frequency plane further suppresses correlations at the expense of resolution, Sec.~\ref{sec:coarse-graining}.

Inference on simulated data with narrow-band signals confirms the above conclusions.
For a sinusoid in powerlaw noise whose spectral index drifts in time, the stationary Whittle likelihood fails to recover most of the SNR, and even fails to detect some signals.
Every WDM-domain analysis with different covariance approximations recovers the signal, but with varying accuracy on the noise.
The dominant effect that causes covariance correlations is the wraparound of the transform, which correlates the start and end of data whose noise levels vary across the observation.
Left unmitigated, wraparound biases the noise parameters significantly.
Replacing the edge columns with the exact covariance handles the effect perfectly, excising them costs SNR, and zero-padding them keeps all the data at the price of a small shift in some noise parameters.
For a resolved Galactic binary in LISA's cyclostationary foreground, the analysis band is narrow and the noise varies slowly.
Every WDM likelihood and covariance approximation returns equivalent signal and noise posteriors, with little sensitivity to the time-edge treatment.
The stationary Whittle likelihood, in contrast, recovers the time-averaged noise level, loses SNR, and broadens the posteriors.

Several directions remain for investigation.
Since the error of a diagonal covariance depends on the choice of time-frequency grid, it may be advantageous to tailor the time-frequency grid to each source class, or even individual sources.
LISA's global fit, envisioned as a blocked Gibbs sampler~\cite{Littenberg:2023xpl}, should afford such flexibility with a different grid per block.
However, broad-band analyses, such as modeling a parametrized Galactic foreground or a massive black hole binary, may still require a tri-diagonal covariance or coarse-graining rather than a diagonal approximation.
An alternative remedy at fixed tiling is to partially whiten the data in the frequency domain by the time-averaged spectrum $\bar{S}(f)=T_\mathrm{obs}^{-1}\int_0^{T_\mathrm{obs}}S(t,f)\diff t$ before applying the WDM transform, which reduces the frequency gradients across cells and hence the off-diagonal terms directly~\cite{Cornish:2026tjt,Cornish:2025awt,Yan:2026nch}.
While moving from frequency-domain to time-frequency domain analyses is clearly advantageous, the exact settings that minimize computational cost and approximation-induced error remain subject to investigation.

\acknowledgements

AJ and RR were supported by the NASA LISA Project Office.
KC was supported by NSF Grant PHY-2409001.
Anthropic's Claude agentic AI has been used to make the figures, verify calculations, proofread the text for typos and inconsistencies, produce the software implementation, and suggest wording edits for clarity.
The authors have reviewed all AI output for correctness. 
This work has made use of the \textsc{numpy}~\cite{harris2020array}, \textsc{scipy}~\cite{scipy}, and \textsc{matplotlib}~\cite{matplotlib} Python packages.

\section*{Data availability}
The code for the likelihoods, covariances, and simulations used in this work is available at \url{https://github.com/AaronDJohnson/wdm_models_scripts}~\cite{wdm_models_scripts}.
The Galactic-binary waveform package \texttt{binary\_galactica} is available at \url{https://github.com/AaronDJohnson/binary_galactica}~\cite{binary_galactica}, and the sampler \texttt{impulse\_mcmc} at \url{https://github.com/AaronDJohnson/impulse_mcmc}~\cite{Johnson2025-fn}.

\bibliography{apssamp}

\end{document}